\documentclass[fleqn,usenatbib]{mnras}
\usepackage{newtxtext}

\usepackage[T1]{fontenc}
\usepackage{ae,aecompl}

\usepackage{graphicx}	
\usepackage{amsmath}	
\usepackage{amssymb}	
\usepackage{bm} 
\usepackage{longtable} 

\newcommand{\beq}{\begin{equation}}
\newcommand{\eeq}{\end{equation}}
\newcommand{\bea}{\begin{eqnarray}}
\newcommand{\eea}{\end{eqnarray}}
\newcommand{\req}[1]{equation~(\ref{#1})}
\newcommand{\dkpc}{d_\mathrm{kpc}}
\newcommand{\erm}[1]{\mathrm{e}^{#1}}
\newcommand{\gcc}{\mbox{g~cm$^{-3}$}}
\newcommand{\kB}{k_\mathrm{B}}
\newcommand{\Reff}{R_\mathrm{eff}^\infty}
\newcommand{\rhob}{\rho_\mathrm{b}}
\newcommand{\sSB}{\sigma_\mathrm{SB}}
\newcommand{\Teff}{T_\mathrm{eff}}
\newcommand{\Ts}{T_\mathrm{s}}
\newcommand{\Ti}{\widetilde{T}}
\newcommand{\Lnu}{L_\nu^\infty}
\newcommand{\Ls}{L_\gamma^\infty}
\newcommand{\msun}{\mbox{M\sun}}
\newcommand{\xmm}{\textit{XMM-Newton}}
\newcommand{\chan}{\textit{Chandra}}

\title{Thermal luminosities of cooling neutron stars}

\author[Potekhin et al.]{
A. Y. Potekhin,$^{1}$\thanks{E-mail: palex-spb@yandex.ru}
D. A. Zyuzin,$^{1}$
D. G. Yakovlev,$^{1}$
M. V. Beznogov,$^{2}$
Yu. A. Shibanov$^{1}$
\\
$^{1}$Ioffe Institute, Politekhnicheskaya 26, 
194021 St Petersburg, Russia
\\
$^{2}$Instituto de Astronomia, Universidad Nacional Autonoma de Mexico,
 Mexico, D.F. 04510, Mexico
 }

\date{Accepted 2020 June 24. Received 2020 June 19; in original form
2020 May 17}

\begin{document}
\label{firstpage}
\pagerange{\pageref{firstpage}--\pageref{lastpage}}
\maketitle

\begin{abstract} Ages and thermal luminosities of neutron stars,
inferred from observations, can be interpreted with the aid of the
neutron star cooling theory to gain information on the properties of
superdense matter in neutron-star interiors. 
We present a survey of estimated ages,
surface temperatures and thermal luminosities of middle-aged neutron
stars with relatively weak or moderately strong magnetic fields, which
can be useful for these purposes. 
The catalogue includes results selected from the literature,
supplemented with new results of spectral analysis of a few cooling
neutron stars.
The data are compared with the theory.
We show that overall agreement of
theoretical cooling curves with observations improves substantially for
models where neutron superfluidity in stellar core is weak. 
\end{abstract}


\section{Introduction}

Comparison of the theory of neutron star cooling with  observations can
provide a wealth of important information. Since the cooling depends on
neutron star properties, for example mass, equation of state (EoS),
composition of the stellar core and outer envelopes, magnetic field
strength and configuration, such a comparison can  help to determine
some parameters of neutron stars and  choose among theoretical models of
superdense matter.

Using the growing number of high-quality observations of thermal
radiation of neutron stars requires a catalogue of their basic
observable parameters relevant to cooling -- first of all, ages and
thermal luminosities. Some compilations have been done in the past,
embedded in reviews of the neutron star cooling  theory (e.g.,
\citealt{YakovlevPethick04,Page09,Tsuruta09,PPP15} and references
therein) or in research papers
\citep[e.g.,][]{Page_04,Zhu_ea11,Vigano_13,BeznogovYakovlev15}. Here we
revisit the collection of observational data on  cooling neutron stars
from the literature, including new results and reanalysing some of  the
archival data. We compose an updated catalogue of observational
estimates of the basic properties of thermally emitting, cooling
isolated neutron stars (INSs). We also present examples of such a
comparison, taking some recent progress of the neutron star theory into
account.

In Sect.~\ref{sect:theory} we sketch the basic concepts of the
neutron star cooling theory. In Sect.~\ref{sec:cooldat} we present the
brief summary of modern observations of thermally emitting, middle-aged
INSs and list their estimated parameters that can be useful for
theoretical interpretation of the data. In Sect.~\ref{sect:cooling} we 
discuss some examples of the comparison of theory and observations.  We
summarize in Sect.~\ref{sect:concl}.

\section{Theoretical background}
\label{sect:theory}

The cooling theory requires constructing the models of
neutron  star structure and thermal evolution. This task is unique in 
the complexity of underlying physics (partially uncertain), ranges of
temperatures and densities, and physical processes  involved.
Detailed reviews were given, e.g., by
\citet*{YakovlevPethick04,Page09,Tsuruta09,PPP15}. 
In this section we outline the basic concepts of this theory.

\subsection{Hydrostatic and thermal structure and evolution}
 
Neutron stars are relativistic objects.  The importance of  General 
Relativity effects is characterized by the compactness ratio
$r_\mathrm{g}/R$, where $R$ is the stellar radius,
$r_\mathrm{g}=2GM/c^2$  is the Schwarzschild radius, $M$ is the
gravitational stellar mass, $G$ is the Newtonian gravitational constant,
and $c$ is the speed of light. Typically, $r_\mathrm{g}/R \sim$
0.2\,--\,0.5 for neutron stars, while $r_\mathrm{g}/R \ll 10^{-3}$ for
all other stars.

The mass density $\rho$ ranges
from  small values in the atmospheres to $\rho \gtrsim 10^{15}$ \gcc{}
at the center of a sufficiently massive neutron star \citep*[e.g.,][]{HPY07}.  
A thin atmosphere covers the ocean, filled with a Coulomb
liquid of atomic nuclei and degenerate electrons, which lies on
the crust, where
the nuclei are arranged in a crystalline lattice.  In addition to the
nuclei and electrons, there are quasi-free
``dripped'' neutrons in the inner crust, at $4.3\times10^{11}~\gcc
< \rho \lesssim 10^{14}$ \gcc. 
At $\rho$ around $10^{14}$ \gcc, the clusters of nucleons
may take exotic non-spherical shapes,
forming a liquid-crystal mantle of the star.
At still higher densities, there is  a 
 neutron-rich liquid core. The central part 
of sufficiently massive neutron
stars (at
$\rho\gtrsim7\times10^{14}$ \gcc) has rather unknown composition and
EoS,  which remains a fundamental
physical problem. Nucleons (and other
baryons if available)  in neutron star interiors can be superfluid.
Neutron stars can  be fast rotators (with spin periods $\mathcal{P}$
down to $\sim1$ ms) and possess strong magnetic fields (up to
$B\gtrsim10^{15}$~G, with typical
values $B\sim10^{12}$~G).

Neutron stars cool down mainly via neutrino emission from their
cores and photon emission from their atmospheres.
While cooling they lose their thermal energy. They may also be 
reheated by various mechanisms \citep[e.g.,][]{GonzalezReisenegger10}. 

Accurate modelling of neutron star cooling with account of all possible
effects is complicated.  The cooling can be affected by 
presence of free hyperons (e.g., \citealt{Grigorian_18,Raduta_19} and
references therein) or deconfined quarks (\citealt{Wei_20} and
references therein), by emission of axions (e.g., \citealt{Sedrakian_19}
and references therein), pion or kaon condensation (see \citealt{YKGH01}
for review and references) and other effects. For simplicity we assume
that the neutron-star matter contains neutrons, protons, electrons and
muons, without hyperons or exotic matter. Also, we restrict ourselves to
cooling INSs with not too strong magnetic fields $B \lesssim 10^{14}$ G
and not too fast rotation ($\mathcal{P}\gtrsim  10$ ms), neglecting 
reheating. Then to a good approximation the internal stellar structure
is spherically symmetric. It is also reasonable to expect that the
temperature distribution is spherically symmetric at sufficiently high
densities. In the region of the star where these assumptions are
acceptable, the mechanical structure and temperature distribution are
determined by a set of differential equations  \citep{Richardson_82},
which involve only one spatial  coordinate, circumferential radius $r$.
To solve this set of equations, one needs an EoS, which relates pressure
$P$ to mass density $\rho$ and temperature $T$, and a boundary
condition, which relates thermal flux to temperature at  an outer
spherical surface. As a rule, the outer boundary is chosen at such
density $\rhob$ that  outer layers at $\rho<\rhob$ can be treated as
quasi-stationary and the matter at $\rho>\rhob$ is strongly degenerate.
Then the EoS at $\rho>\rhob$ is nearly barotropic ($P$ depends only on
$\rho$ but not on $T$), and the solution of the cooling problem can be
greatly simplified: one first solves the hydrostatic equilibrium
equations, once and forever, and then uses the stationary hydrostatic
structure in the heat transport and thermal balance equations. Many
successful neutron star cooling codes work in this way (e.g.,
\citealt{GYP01,NSCool}). There are also more complicated codes for
simulations of neutron-star thermal evolution beyond the approximation
of barotropic EoS (e.g., \citealt{Richardson_82,PC18};
\citealt*{BeznogovPR20}); they are needed to study hot neutron stars or
neutron stars with rapid variations of thermal emission. More complex
codes that take into account departures from spherical symmetry are
required for studies of rapidly rotating or ultra-magnetized neutron
stars (e.g., \citealt{Chatterjee_15,PonsVigano19} and references
therein).

\subsection{Heat-blanketing envelopes}
\label{sect:blanket}

The above-mentioned quasi-stationary layer is called
\emph{heat-blanketing envelope}. Usually, following \citet*{GPE83}, one
sets its bottom at the mass density $\rhob = 10^{10}$ g cm$^{-3}$ (a few
hundred meters under the surface).  \citet{PC18} have explicitly
demonstrated that this choice provides a good accuracy to the cooling
simulations of INSs at ages $t\gtrsim1$~yr. However, the envelope can be
chosen thinner or thicker depending on the problem under consideration.

The composition of the
heat blankets is generally unknown. In proto-neutron stars,
thermonuclear burning produces
 the iron group composition near the surface.
  However, light elements may be supplied
by the fallback of matter onto the stellar surface, by accretion from
interstellar medium or a companion star and by other processes.

Different chemical elements are separated by rapid sedimentation caused
by the strong gravity of neutron stars \citep*{HameuryHB83}, although
thermonuclear processes in the envelopes of accreting neutron stars can
instantaneously create  complex mixtures  (see, e.g.,
\citealt{Meisel_18} for review and references). \citet*{PCY97} studied
the blanketing envelopes composed, from surface to bottom, of hydrogen,
helium, carbon, and iron shells, assuming perfect stratification. More
scrupulous envelope treatments include smearing the interfaces between
the shells by diffusive mixing of different ions
\citep*{DeBlasio00,CB03,CB10,BPY16}.

In strong magnetic fields, the heat conduction in the envelopes is
anisotropic, so that the temperature varies over the
surface
\citep{Greenstein}. However, the total thermal photon luminosity
$L_\gamma$ is only weakly affected by this anisotropy \citep{PY01,PYCG03}, except for
superstrong fields $B\gtrsim10^{14}$~G, which appreciably increase the
overall heat-transparency of the envelope \citep*{PUC,PC18}. 

\subsection{Observables}

Since the distribution of the surface temperature $\Ts$ over the
neutron-star surface can be non-uniform, it is
convenient to introduce the overall effective temperature of the star,
$\Teff$, defined by
\beq
   4 \pi \sSB R^2 \Teff^4 =
   L_\gamma= 
 \int F_\mathrm{s} \,\mathrm{d}\Sigma =
\sSB \int \Ts^4 \,\mathrm{d}\Sigma ,
\label{L}
\eeq
where
$F_\mathrm{s}$ is the local flux density at the radiative surface
and
 $\mathrm{d}\Sigma$ is the surface element.
The quantities $\Ts$, $\Teff$, and $L_\gamma$
refer to a local reference frame at the neutron-star surface.
The quantities detected by a distant observer are redshifted
\citep[e.g.,][]{Thorne77},
\begin{eqnarray}
&&   L_\gamma^\infty = L_\gamma (1 - r_\mathrm{g}/R) =
     4 \pi \sigma_\mathrm{SB} (T_\mathrm{eff}^\infty)^{4} R_\infty^2,
\label{srt-L_gamma_infty}\\
&&   T_\mathrm{eff}^\infty = T_\mathrm{eff} \, \sqrt{1 - r_\mathrm{g}/R}, \quad
     R_\infty = R/ \sqrt{1 - r_\mathrm{g}/R}.
\label{therm-T_s_infty}
\end{eqnarray}
%

\subsection{Basic regulators of neutron star cooling}
\label{sect:regulators}

After a newly born neutron star has passed the initial stage of
internal thermal relaxation, which lasts $\sim10^2$~yr (see
\citealt{GYP01}), it is almost isothermal inside, excluding heat
blankets. The isothermality means that the \emph{redshifted} temperature
$\Ti=T(r)\,\erm{\Phi(r)}$ is independent of $r$ (here, $\Phi(r)$ is a
dimensionless metric function; e.g., \citealt*{MisnerTW}). Subsequent
thermal evolution of an INS passes through the neutrino cooling stage
and the photon cooling stage, possibly followed by reheating.  The
time-dependence of $\Ti$ is governed by the heat balance equation
\begin{equation}
   C(\Ti)\,\frac{{\rm d}\Ti}{{\rm d}t}=
   - \Lnu (\Ti)- \Ls (\Ti).
\label{e:isocool}
\end{equation}
Here, $C(\Ti)$ and $\Lnu(\Ti)$ are, respectively, the total heat
capacity and neutrino luminosity of the star, properly integrated over
the  stellar volume (see, e.g., \citealt*{YLH03}).

INSs of age $t \lesssim (0.1-0.3)$ Myr cool mainly via neutrino 
emission ($\Lnu \gg \Ls$), so that $\Ti(t)$ is regulated by the ratio
$\Lnu(\Ti)/C(\Ti)$.  Older INSs cool via thermal emission of photons
from the surface. The neutrino term $L_\nu$ becomes negligible in
\req{e:isocool}. At this stage the cooling is driven by the ratio
$\Ls(\Ti)/C(\Ti)$. At each stage, the dependence $\Ls(\Ti)$, which is
provided by a heat blanket model, is  needed to obtain the observable
photon luminosity. 
Therefore, using the 
cooling theory, one can test the heat capacity and neutrino luminosity
of the stellar core, $C$ and $\Lnu$, and heat transparency of the
blanketing envelope (determined by composition of plasma as well
as by magnetic field strength and geometry).

Microphysics of the crust does not strongly affect the 
cooling of INSs. It can affect cooling of younger stars (where the crust
is non-isothermal) or stars with very strong fields $B \gtrsim
10^{14}$~G (see \citealt{PC18}).

The functions $C(\Ti)$ and $\Lnu(\Ti)$ are mostly determined by properties of
superdense matter in the core. One can distinguish the effects of two
types.

The effects of first type come from nuclear physics. If the proton
fraction (which is strongly model-dependent) can reach sufficiently high
values in the core, then very fast neutrino cooling via direct Urca
processes can proceed (see, e.g., \citealt{Haensel95}, and references 
therein). Otherwise the neutrino cooling would be much
slower, being mediated mainly by modified Urca processes and neutrino
bremsstrahlung in nucleon collisions.

The effects of second type are regulated by superfluidity of neutrons
and protons (see \citealt{Page_14} for review). In the core, neutron
pairing may occur in triplet state while proton pairing is in
singlet state.  In the inner crust, dripped neutrons undergo
singlet-state pairing. The key theoretical ingredients are critical
temperatures $T_\mathrm{cn}$ and $T_\mathrm{cp}$ for neutron
and proton pairing as functions of $\rho$, which are
related to the so called superfluid gap functions $\Delta(k_\mathrm{F})$,
where $k_\mathrm{F}$ is the baryon Fermi wave-number.
Their calculation is very complicated, 
being affected by numerous delicate effects of in-medium
interaction of nucleons (see,
e.g., \citealt{SedrakianClark} for review). 
Modern calculations of the
$\Delta(k_\mathrm{F})$ profiles converge to similar results
for the $^1S_0$ pairing of neutrons
\citep{MargueronSH08,GIPSF,Ding_16}, which is at work in the crust, 
but $\Delta(k_\mathrm{F})$ for the singlet pairing of protons
and for triplet pairing of neutrons in the core 
vary by orders of magnitude
depending on employed theoretical model.

Strong pairing of nucleons suppresses their heat capacity and neutrino
cooling reactions involving these nucleons (see \citealt{YKGH01} for
review and practical analytic expressions). When the core temperature
becomes much lower than a critical temperature, the suppression can be
exponential (almost complete).  The  suppression of neutrino reactions
slows the cooling down, while the suppression of the heat capacity
accelerates it. The former effect dominates at the neutrino cooling
stage, and the latter becomes most important when the star cools mainly
via photon emission.  Besides, superfluidity can accelerate the cooling
via an additional moderately strong neutrino emission due to Cooper
pairing of neutrons. This emission is substantial in the triplet
pairing channel when temperature is slightly below $T_\mathrm{cn}$
(\citealt{Leinson10} and references therein). Therefore, even the
direction of superfluidity effects can be  different, but all this
physics is compressed in the two functions, $C(\Ti)$ and $\Lnu(\Ti)$.
Inferring these functions from the data can give useful information on
internal structure of neutron stars.


\section{Observations of cooling neutron stars}
\label{sec:cooldat}

\subsection{General remarks}
\label{sect:gen}

Only a small fraction of the observed neutron stars is suitable for
comparison with the theory of cooling. A vast majority of
neutron stars demonstrate intense emission of non-thermal origin. 
Neutron
stars in binary systems are usually surrounded by an accretion disk,
whose emission is orders of magnitude more powerful than the thermal
emission from the neutron star surface. 
Non-thermal emission of INSs can be produced 
by pulsar activity or by other processes in the
magnetosphere of the star. 
A careful analysis is required to extract the
thermal component of the observed spectrum.

On the other hand, very old neutron stars,
including almost all millisecond pulsars, have already lost their
initial heat. Their thermal luminosity is very low, and even if
detected, it can be produced by reheating
\citep*[e.g.][]{GonzalezReisenegger10,GJPR15,YanagiNH20}
or by hot polar caps heated by return currents 
\citep[e.g.,][]{TimokhinArons13,Salmi_20}.

Another problem is that the age of neutron stars is rarely known with
good accuracy. Thus, the ``passive  cooling'' theory, which neglects
reheating, can only be  tested against observations of not too old
INSs with estimated ages. 

In order to compare neutron-star observations with the cooling theory, 
one may use either surface temperatures $T_\mathrm{s}$ or photon
luminosities $L_\gamma$.  Both can be evaluated by spectral analysis,
but it is usually difficult to determine them accurately. In empty flat
space, the observed flux is inversely proportional to squared distance,
while the shape of the spectrum is distance-independent. At the first
glance, temperature measurements should be preferred, since the distance
is often poorly known. In reality, however, absorption by interstellar
medium appreciably affects both the luminosity and the spectral shape.
At typical surface temperatures of cooling neutron stars
$T_\mathrm{s}\sim\mbox{ a few}\times(10^5-10^6)$~K, thermal flux is
emitted mostly in soft X-rays, but a substantial part of the thermal
spectrum lies in the extreme ultraviolet (EUV) range, which is
inaccessible to observations because of the strong interstellar
absorption. For these reasons, a complete recovery of the spectral shape
is a problem even for a neutron star with purely thermal emission.  For
nearby stars detected in the optical-UV, additional constraints  can be
obtained from the  Rayleigh-Jeans tail of the thermal spectrum 
\citep[e.g.,][]{Shibanov_05,KargaltsevPavlov07,Kaplan_11}.

The analysis of the spectrum
can be performed  under different hypotheses on the thermal and
non-thermal components. Both temperature and luminosity estimates depend
on the choice of the emission model. The simplest
model is the blackbody (BB) spectrum. More physically motivated models rely
on computations of radiative transfer in neutron-star atmospheres or
formation of thermal spectrum at a condensed surface (see, e.g.,
\citealt{Potekhin14} for review). Often more than one model can fit
equally well the data, without any clear, physically motivated
preference. Estimates of the surface temperature,
being extracted from observational data by using different models, may
vary by a factor of a few. As a rule, they
anticorrelate with radius of an equivalent emitting sphere for a
distant observer, $\Reff$. Often the model that yields a higher
estimate of temperature gives a lower luminosity (mainly because of
a smaller bolometric correction). Since the neutron star mass $M$
and physical radius $R$ are rarely known with good accuracy,
estimates of redshifted temperature
$T^\infty$ and estimates of redshifted
photon luminosity $L^\infty$ are
usually more robust than non-redshifted ones.

If data contain few photons and/or strong absorption features, 
the temperature may be poorly constrained by the fit, adding a large
statistical error to the systematic one.  In such cases $L_\gamma$ can
be determined more accurately than $T_\mathrm{s}$, especially if the
distance to the source is well known. On the other hand, for bright
sources with poorly known distances, the observed temperature can be
constrained better than the luminosity within a
fixed spectral model, although different models still 
give different temperature estimates.

Even an
accurate estimate of surface temperature may be insufficient for comparison with the
cooling theory without a reliable estimate of the luminosity. Indeed, it
is $L_\gamma^\infty$ that enters \req{e:isocool}. In the ideal case where
$T_\mathrm{s}$ is the same at any point on the surface, it equals $\Teff$
in Eq.~(\ref{srt-L_gamma_infty}), where, for
consistency, one should use the same radius $R$
that is employed in
the cooling simulation being compared with the observations. Then the
knowledge of $T_\mathrm{s}$ is equivalent to the knowledge of
$L_\gamma$. However, most neutron stars (including all neutron stars with
strong magnetic fields) have anisotropic distribution of
temperature over the surface (Sect.~\ref{sect:blanket}).  Then a single
temperature extracted from observations 
may be biased upwards, because hotter regions
contribute more to the detected flux. Accordingly, the effective
emitting area is often smaller than $4\pi R^2$, so that  $\Reff <
R_\infty$. Since the energy balance includes the total photon flux, it
would be inconsistent to compare theoretical surface temperature, calculated 
for the entire emitting surface, with the observational estimate
that implies that only a part of the surface emits
thermal  photons. To restore consistency,
$\Ls$ in \req{e:isocool} ought to be corrected by
using an appropriate integration area in \req{L}.

More sophisticated models allow for a non-uniform temperature
distribution. For example, observed spectra are often fitted with
models, which include two, or sometimes more, thermal components with
different $T^\infty$ and  $\Reff$. Usually the hottest component has a
small effective radius ($\Reff\ll R_\infty$), which may correspond, for
example, to a hot spot on the surface, such as a polar cap of a pulsar
heated up by return currents. The hot spots provided by such models usually
give minor contribution to the total luminosity. In some cases, however,
different fit components have comparable effective radii and give
comparable contributions to $L^\infty$. Such a step-like temperature
distribution may be an artifact, caused by inadequacy of the assumed
spectral model, but it also may mimic a real variation of temperature
over the surface.

As will be seen from the comments on observational data in
Section~\ref{sec:commentsources}, in some cases the temperature and in
other cases the luminosity is better constrained by the data. 
Sometimes, estimates of both the luminosity and the temperature together
help to constrain the parameters of a given  neutron star. Thus it is
advisable to take into account the totality of observational data for an
analysis of every individual source under study.

\subsection{Description of tables}
\label{sect:descriptables}

Table~\ref{tab:CoolList} gives basic information on middle-aged INSs with registered or
reliably constrained thermal radiation that are suitable for comparison
with the cooling theory. The key properties of these
neutron stars for such a comparison are given in Table~\ref{tab:AgeL}.

In Table~\ref{tab:CoolList}, we list the principal identifier of a
neutron star, its name or the name of related supernova remnant (SNR),
nebula, or stellar association, its spin period $\mathcal{P}$,
characteristic dipole field $B_\mathrm{dip}$, and distance $d$. 
Question marks indicate uncertain associations, which are not used in
Table~\ref{tab:AgeL}. The characteristic field is defined by the
standard formula \citep[e.g.,][]{ManchesterTaylor}
$B_\mathrm{dip}=3.2\times 10^{19}
\,\sqrt{\mathcal{P}\dot{\mathcal{P}}}$~G, where $\mathcal{P}$ is in
seconds. It gives the field at the magnetic equator of the rotating
orthogonal non-relativistic magnetic dipole in vacuo with
\emph{canonical neutron star} parameters: mass $M=1.4\,\msun$, radius
$R=10$~km and moment of inertia $I=10^{45}$ g~cm$^2$.

Table~\ref{tab:AgeL} gives the characteristic age
\citep[e.g.,][]{ManchesterTaylor} $t_\mathrm{c} =
\mathcal{P}/(2\dot{\mathcal{P}})$, the age $t_*$ estimated independently
of timing, the redshifted bolometric luminosity $L^\infty$ and
redshifted surface temperature in energy units ${\kB}T^\infty$. An
estimate $t_*$ can be based on proper motion of the star, on physical
properties of the associated SNR or surrounding nebula, or, in a few
cases, on historical supernova dates (in these cases $t_*$ gives
the time of reported spectral measurements). The characteristic ages are
usually treated as upper limits. These limits are rather loose:  the
true ages can sometimes be a factor of 2\,--\,3 longer than
$t_\mathrm{c}$ (for example, for pulsars Vela and J1119$-$6127; see
Sect.~\ref{sec:commentsources}). Nevertheless, in most cases 
$t_* < t_\mathrm{c}$ (cf.{} Table~\ref{tab:AgeL}), sometimes even
by orders of magnitude (as for young neutron stars RX
J0822.0$-$4300, 1E 1207.4$-$5209 and CXOU J185238.6+004020).

If a spectral fit includes both thermal and
non-thermal components, only the thermal component of the total
luminosity is given. If two or more thermal components are included in
the spectral fit, we list the temperature for the cooler one and, over
the slash, the hotter one,  provided the latter contributes
substantially (by more than 15\%) to the total flux, and we do the same for
the effective radius in the comments. For example,
`${\kB}T^\infty=154\pm4/319^{+13}_{-12}$ eV,
$\Reff=2.21^{+0.08}_{-0.07}/0.37\pm0.04$~km' means that one spectral
component has been fitted with ${\kB}T^\infty=154\pm4$ eV and
$\Reff=2.21^{+0.08}_{-0.07}$ and the other component with 
${\kB}T^\infty=319^{+13}_{-12}$ eV and $\Reff=0.37\pm0.04$~km, so that
each component provides more than 15\% of the total luminosity. In such
cases, the two components may mimic a real inhomogeneity of temperature
distribution over the surface. If the energy contribution of the
hotter component is small, it is not listed, because in this case it has
a small $\Reff$ and probably represents a hot polar cap heated by
reverse currents from the magnetosphere.

The last column of each table indicates references. The reference
numbers are marked by letters: `d' for distance, `a' for the estimated
age $t_*$, and `s' for spectral analysis. The quantities defined by
timing ($\mathcal{P}$, $t_\mathrm{c}$ and $B_\mathrm{dip}$) 
mostly rely on the ATNF Pulsar Catalogue
\citep{ATNF},\footnote{\url{https://www.atnf.csiro.au/research/pulsar/psrcat/}}
to which the reader is addressed for primary references; otherwise we
give a reference marked by letter `t' in Table~\ref{tab:CoolList}. For
pulsars with measured parallaxes, we mostly quote distances corrected
for the Lutz-Kelker bias from \citet{Verbiest_ea12}, where the reader
can find references to original measurements. Some explanations about
the  listed estimates for each neutron star are given in the comments
below.

The objects in the tables are grouped in several classes. Within each
class, they are sorted by their mean equinox.

The first class includes central compact objects (CCOs) in SNRs and
other thermally emitting isolated neutron stars (TINSs) without a clear
SNR association, which mostly show soft X-ray thermal-like radiation and
are not very strongly magnetized ($B_\mathrm{dip}$ is below
$5\times10^{11}$~G or non-determined). 

The second and third classes are composed, respectively, of ordinary
rotation-powered pulsars with moderately strong magnetic fields
$B\sim(10^{12}-10^{13})$~G  and high-B pulsars with
$B\sim(10^{13}-10^{14})$~G. Such magnetic fields make the temperature
distribution over the surface strongly non-uniform, but they only weakly
affect thermal luminosity $L_\gamma$ (see Sect.~\ref{sect:blanket}). The
first pulsars with
clearly  measured thermal X-ray spectra, PSR B0656+14, B1055$-$52, and
Geminga, were nicknamed the Three Musketeers by \citet{BeckerTrumper97}.

The fourth class consists of the X-ray emitting neutron stars (XINS). 
These seven INSs, nicknamed \emph{The Magnificent Seven}, were
discovered in the \textit{ROSAT All Sky Survey} (see
\citealt{Haberl07,Turolla09,KvK09}, for reviews). Their soft X-ray
radiation can be mostly thermal, but unlike CCOs and TINSs, they have
very strong magnetic fields, similar to those of the high-B pulsars.

Next, we present a bunch of not too old pulsars with upper bounds on
their thermal luminosity,  which may be useful for constraining
cooling predictions.

Finally we list a group of pulsars whose effective thermally emitting
areas are very small ($\Reff \lesssim0.5$~km), suggesting that their
thermal radiation may originate mostly in the polar caps. These
estimates may be useful, because the registered hot-cap luminosity can
be regarded as an upper bound on the total thermal luminosity. However,
these upper bounds  are not absolute, because a soft thermal emission
from the entire surface might escape detection. For example,
\citet{RigoselliMereghetti18}  derived $3\sigma$ upper bounds of
$3.2\times10^{28}$ erg s$^{-1}$ and $2.4\times10^{29}$ erg s$^{-1}$ on
the bolometric luminosities of  a possible thermal component with   $\kB
T^\infty$ between 50 eV and 2 keV for the pulsars B0628$-$28 and
B0919+06, respectively.  Meanwhile, for a neutron star with typical
$R_\infty\sim15$ km, thermal radiation with $\kB T^\infty\sim25$ eV
(below the formal applicability range of these bounds) would correspond
to $L^\infty\sim10^{31}$ erg s$^{-1}$ (well above these limits). On the
other hand, there are a number of  thermal-like spectra of pulsars,
which could be interpreted as produced by hot spots while fitted with 
the blackbody spectrum, because of rather low inferred $\Reff$ and high
$T^\infty$, whereas the interpretation by a thermal emission from the
entire surface becomes possible when an atmosphere model is applied.

The ages and observational estimates of thermal
luminosity or surface temperature from Table~\ref{tab:AgeL} are compared with theoretical
cooling curves in Sect.~\ref{sect:cooling}.

\subsection{Comments on individual objects}
\label{sec:commentsources}

The objects in the list below are numbered in the same order as in
Tables~\ref{tab:CoolList} and~\ref{tab:AgeL}.

\subsubsection{Weakly magnetized thermally emitting neutron stars}

1. \emph{1E 0102.2$-$7219} belongs to a SNR, which was revealed in
X-rays by the \textit{Einstein} observatory. Its location in  the Small
Magellanic Cloud fixes its distance to $d=62\pm2$ kpc
(\citealt{SMCdistance} and references therein). Its age was estimated in
a number of papers (see \citealt{Xi_19} and references therein).
\citet{Vogt_18} discovered the respective CCO in the data of the
\textit{Chandra} X-ray observatory.  \citet*{HebbarHH20} confirmed this
detection and performed a spectral analysis using archival
\textit{Chandra} observations. We adopt their preferred result
(${T}_\mathrm{s}=3.0^{+0.5}_{-0.4}$ MK,
$\Reff/R_\infty=0.5^{+0.5}_{-0.2}$, 
$N_\mathrm{H}=9^{=12}_{-7}\times10^{21}$ cm$^{-2}$, $L^\infty =
1.1^{+1.6}_{-0.5}\times10^{34}$ erg s$^{-1}$), which was obtained with
the \textsc{nsmax} spectral model\footnote{The named atmosphere models
(\textsc{nsmax} by \citealt*{HoPC08}; \textsc{nsa} by
\citealt*{ZavlinPS96}; \textsc{nsmaxg} by \citealt{Ho14}; \textsc{nsx}
by \citealt{HoHeinke09}; \textsc{carbatm} by \citealt{Suleimanov_14};
\textsc{tbabs} by \citealt*{WilmsAMC00}) are from the XSPEC package
\citep{XSPEC} at \url{https://heasarc.gsfc.nasa.gov/docs/xanadu/xspec/}}
of an atmosphere composed of partially ionized carbon at magnetic field
$B=10^{12}$~G on a neutron star with fixed $M=1.4\,\msun$ and $R=12$ km
with free normalization of the point source and background fluxes. This
model provides the only satisfactory one-component spectral fit. Other
statistically acceptable fits include multiple spectral components and
suggest temperatures and luminosities varying in substantially wider
ranges. Besides, the authors warn that  the high contribution from the
background makes the statistical confidence of different fits somewhat
uncertain.

2. \emph{RX J0822.0$-$4300 (PSR J0821$-$4300)} is a CCO in the SNR
Puppis A (G260.4$-$03.4). Its age was estimated from the proper motion
measurements \citep{Becker_ea12}. The two blackbody (2BB) fit to the
spectrum at fixed distance $d=2.2$ kpc  gives 
${\kB}T^\infty=265\pm15/455\pm20$ eV with  $\Reff$  varying from
2.27/0.53 km to 2.04/0.65 km between the ``soft'' and ``hard'' phases of
the pulse profile \citep{DeLuca_ea12}.  The sum of the cooler and hotter
components thus varies from  $L^\infty=(4.8^{+1.1}_{-0.9})\times10^{33}$
erg s$^{-1}$ in the ``soft'' phase to $L^\infty=5.0^{+1.1}_{-0.9}\times
10^{33}$ erg s$^{-1}$ in the ``hard'' phase of the pulse.

3. \emph{CXOU J085201.4$-$461753} is a CCO in the SNR Vela Jr.
(G266.2$-$1.2). Its age and distance were estimated from an expansion
rate and a hydrodynamic analysis \citep{Allen2015-velajr}. Using the
\textsc{nsx} model of the carbon atmosphere, \citet{Danilenko_15} showed
that the X-ray spectrum of this CCO  agrees with thermal radiation of a
neutron star at $T^{\infty} = 1.3$ MK. An alternative BB fit implies a
small emitting area, which is difficult to agree with the tight
constraint on the pulsed fraction $<3\%$.

4. \emph{2XMM J104608.7$-$594306}. This neutron star, possibly in the
Homunculus nebula around $\eta$ Carina, has purely thermal X-ray
spectrum. Its age estimate is based on a proposed association of a
nearby star with its progenitor \citep{Pires_ea15}. Having performed an
analysis based on different spectral models, \citet{Pires_ea15}
concluded that the spectrum can only be well fitted with including at
least  one absorption line at energy 1.35 keV. If interpreted as a
redshifted electron cyclotron line, it implies
$B\sim1.5\times10^{11}$~G.  In this case, 
${T}_\mathrm{s}=(6-10)\times10^5$~K and the X-ray luminosity 
${L}_\mathrm{X}=(1.1-7.4)\times10^{32}$ erg s $^{-1}$. Taking into
account the gravitational redshift and the bolometric correction, this
translates into $\kB T^\infty=40-70$ eV and
$L^\infty=(0.8-6)\times10^{32}$ erg s $^{-1}$.

5. \emph{1E 1207.4$-$5209 (PSR J1210$-$5209)} is a CCO in the SNR
G296.5+10.0. The estimate and uncertainties of the luminosity are
obtained by comparison of three fitting models at fixed $d=2$ kpc in
Table~1 of \citet{Mereghetti_ea02}. The corresponding effective radius
is $\Reff=2.1\pm0.3$~km. \citet{Roger_ea88} give $t_*\sim7$ kyr
with an uncertainty up to a factor of 3.

6. \emph{1RXS J141256.0+792204 (PSR J1412+7922, RX J1412.9+7922)}. This
enigmatic TINS was initially considered as a possible
addition to the `Magnificent Seven' and
was dubbed `Calvera'  \citep*{RutledgeFS08}. However,
subsequent observations suggest that its properties are closer to the
CCOs. \citet*{HalpernBG13} characterized Calvera as an `orphaned CCO,'
whose magnetic field is emerging through supernova debris. They tried to
estimate the distance and age of Calvera and concluded that both remain
uncertain by an order of magnitude. \citet{Zane_ea11} performed the
first detailed analysis of its spectrum. Among numerous fits tried by
the authors, only those with the interstellar hydrogen column density
$N_\mathrm{H}$ fixed to the Galactic value $2.7\times10^{20}$ cm$^{-2}$
gave an acceptable reduced chi-square statistic
$\chi_\nu^2$ and a reasonable distance (other fits
gave distances down to 175 pc, too large $N_\mathrm{H}$ and tiny
emitting area). A fit with a two-temperature atmosphere model (2NSA)
gave ${\kB}T^\infty=67^{+7}_{-12}$ eV / $150^{+12}_{-20}$ eV, with
radius of the cold component fixed at the realistic value $R=12$~km.
This fit gave $L^\infty\sim6\times10^{32}$ erg~s$^{-1}$ at
$d\approx1.55$ kpc. Other acceptable fits indicated $d$ in the range
from $\sim1$ kpc to $\approx2.25$ kpc. The values of the unabsorbed flux
for various statistically acceptable fits implied $L^\infty$ ranging
from $5.5\times10^{32}$ erg~s$^{-1}$ to $9.2\times10^{32}$ erg~s$^{-1}$,
assuming $d=2$ kpc. \citet{Shibanov_16} fitted the \textit{XMM-Newton}
and \textit{Chandra} data using \textsc{nsmax} models 
of
magnetized hydrogen atmospheres covering the entire surface, assuming a
centered dipole field and the ensuing temperature distribution. They
found that an additional absorption feature at energy
$E_\mathrm{abs}=740\pm30$ eV improved the fit. In this case, the
best-fit model yields $T^\infty=1.18\pm0.05$ MK. The resulting
normalization  (the effective emitting radius) can be reconciled with 
 $R=12$~km  (assumed in the \textsc{nsmax}
model) at $d\sim4$ kpc. The corresponding
luminosity is $L^\infty=(3.0\pm0.5)\times10^{33}$ erg~s$^{-1}$.
Other fits, performed for different data sets with or without involving
the absorption, are only slightly worse statistically. They yield
$T^\infty$ from $0.79\pm0.03$~MK  to $1.34^{+0.01}_{-0.02}$~MK, 
corresponding to $L^\infty\sim5\times(10^{32}-10^{33})$ erg~s$^{-1}$. 
\citet{Bogdanov_ea19} have performed an analysis of the spectrum obtained
with \textit{NICER}. They fixed $d=2$ kpc and obtained $\kB
T^\infty=205\pm3$ eV, $\Reff=1.01^{+0.04}_{-0.03}$~km ($L^\infty =
2.33^{+0.29}_{-0.14}\times10^{32}$ erg~s$^{-1}$) for a blackbody plus
power-law (BB+PL) fit with an absorption at $E_\mathrm{abs}=780\pm20$
eV.  An alternative fit including  a Gaussian absorption at
$E_\mathrm{abs}=760\pm10$ eV  and two blackbody components (the G2BB
fit) gives $\kB T^\infty=154\pm4/319^{+13}_{-12}$ eV,
$\Reff=2.21^{+0.08}_{-0.07}/0.37\pm0.04$~km, totalling to $L^\infty=
(5.4\pm0.6)\times10^{32}$ erg~s$^{-1}$.  The ranges of $L^\infty$ and
$T^\infty$ in Table~\ref{tab:AgeL} cover the results obtained by
\citet{Bogdanov_ea19} and by \citet{Shibanov_16} for the different
instruments.

7. \emph{CXOU J160103.1$-$513353} is a CCO in the SNR G330.2+1.0.  
Proper motion of the SNR fragments indicates an age of
0.8\,--\,1 kyr, if one neglects deceleration;
because of deceleration the actual age should be smaller
 \citep{Borkowski2018-age-cco-snrg330,Williams_18}.
\citet{McClureGriffiths_01} derived the minimum distance to this SNR of
$d=4.9\pm0.3$ kpc from the observed velocities in the absorption spectrum;
a hypothetical maximum distance is $d<9.9$ kpc. The X-ray spectrum of
the CCO is  well described with either single-component carbon or
two-component  hydrogen atmosphere models
\citep{Doroshenko2018-cco-snrg330}.  In the latter case,  the observed
spectrum is dominated by the emission from a hot component with a
temperature $\sim3.9$ MK, corresponding to  the emission from a hot spot
occupying 1\% of the stellar surface, assuming a neutron star with $M =
1.5\,\msun$ and $R=12$ km at $d\sim5$ kpc. 
The carbon atmosphere model yields more plausible results.
Using absorbed  \textsc{carbatm} model  with fixed $d=4.9$
kpc,  $M = 1.5 \msun$ and $R = 12$ km,
\citet{Doroshenko2018-cco-snrg330} obtained non-redshifted
temperature $\Ts = 1.73\pm0.01$ MK. 

8. \emph{1WGA J1713.4$-$3949} is a compact X-ray source, which was
suggested to be a neutron star associated with the SNR G347.3$-$0.5
\citep{Slane_99,Lazendic_03,CassamChenai_04}. Alternatively, 1WGA
J1713.4$-$3949 may be a background extra-galactic source
\citep{Slane_99}. Radio pulsar PSR J1713$-$3945 at the center of the SNR
G347.3$-$0.5 is unrelated to 1WGA J1713.4$-$3949 \citep{Lazendic_03} and
thus could be considered as an alternative association candidate.
However, the large characteristic age $t_\mathrm{c}=1.1$~Myr of PSR
J1713$-$3945 and its dispersion-measure distance estimate of 4.3~kpc
make it unlikely to be associated  with the SNR G347.3$-$0.5
\citep{Lazendic_03}. Location of the SNR G347.3$-$0.5 probably coincides
with the historical supernova SN 393 \citep{WangQC97}, giving the age
around $1608$~yr at the time of  the \xmm\ spectral observations. The
SNR distance is $d=1.3\pm0.4$ kpc, as argued by \citet{CassamChenai_04}
and corroborated, e.g., by  \citet{Moriguchi_05} and \citet{Fukui_12}.
The distance $\sim1$ kpc is also compatible with $t_*\sim1.6$ kyr
\citep{Maxted_13}. A BB+PL fit to the X-ray spectrum ($\chi_\nu^2=1.06$)
gives ${\kB}T^\infty=400\pm20$~eV, $\Reff=0.35\dkpc$ and 
$L^\infty=(3.2\pm0.4)\times10^{32}\dkpc^2$ erg s$^{-1}$, where
$\dkpc\equiv d/\mbox{1~kpc}$ \citep{CassamChenai_04}.  This is
compatible with the hot spot emission, but the lack of pulsations casts
doubt on  such interpretation. The 2BB fit is slightly better
($\chi_\nu^2=1.03$); it gives ${\kB}T^\infty=320^{+20}_{-30}$~eV,
$L^\infty=(4.4\pm0.4)\times10^{32}\dkpc^2$ erg s$^{-1}$ for the cold
component and ${\kB}T^\infty=570^{+70}_{-50}$~eV,
$L^\infty=1.7^{+0.6}_{-0.3}\times10^{32}\dkpc^2$ erg s$^{-1}$ for the
hot component, but again with small effective radii $\sim0.1-0.6$ km. 

Our analysis of archival \xmm\ EPIC-pn data  (OBsID 0740830201), using
the \textsc{tbabs} $\times$ \textsc{nsx} model with fixed  $M = 1.4
\msun$ and $R = 13$ km, gives $T^{\infty} = 1.60\pm0.05$ MK,
$L^\infty=(1.1\pm0.1)\times10^{34}$ erg s$^{-1}$ and 
$d=3.20^{+0.28}_{-0.26}$ kpc (at 95\% confidence), inconsistent with the
distance to G347.3$-$0.5. The same fit with fixed $d=1$ kpc requires
normalization of 0.2 (i.e., $\Reff\approx7$ km), which gives
$L^\infty\sim2\times10^{33}$ erg s$^{-1}$.

9. \emph{XMMU J172054.5$-$372652} is a compact thermal X-ray source,
probably associated with the SNR G350.1$-$0.3, as suggested by 
\citet{Gaensler2008-cco}, who also proposed that the SNR G350.1$-$0.3 
is probably
interacting with a molecular cloud at $d\approx4.5$ kpc.
\citet{Lovchinsky_11} showed that the SNR G350.1$-$0.3 is
in  the free expansion stage and calculated an age of 600\,--\,1200
years.

We used archival \chan\ (OBsID 10102, 14806) data and fitted the
spectrum with the \textsc{tbabs} $\times$ \textsc{nsx} model 
in the range 0.5 -- 6 keV. We fixed
neutron star mass $M = 1.4 \msun$, radius $R = 13$ km and distance
$d=4.5$ kpc. The non-redshifted temperature is
${T}_\mathrm{s}=2.06\pm0.02$ MK 
($T^\infty=1.71\pm0.02$ MK; $L^\infty=(1.51\pm0.07)\times10^{34}$ erg
s$^{-1}$). Letting distance to vary, we obtained
$d=6.1^{+2.6}_{-1.9}$ kpc and 
$T^\infty=1.87^{+0.21}_{-0.20}$ MK (95\% confidence). In each case we got
$\chi_\nu^2=1.03$ (356 and 355 d.o.f.,
respectively). 

10. \emph{XMMU J173203.3$-$344518} is a CCO in the SNR HESS J1731$-$347
(G353.6$-$00.7). For this X-ray source, two likely distance values used
to be considered, $d\sim3.2$ kpc or $4.5$ kpc \citep{Abramowski_ea11}.
The spectral analysis by \citet{Klochkov_etal15} gave an acceptable
neutron star radius $R=12.4^{+0.9}_{-2.2}$~km only for $d\sim3.2$ kpc
and only for the carbon atmosphere model. The analysis suggested the mass
$M=1.55^{+0.28}_{-0.24}$. Later the distance $d\approx3.2$ kpc was
confirmed by \citet{Maxted_ea18}, based on an analysis of
photoabsorption by neutral hydrogen.  \citet{Klochkov_etal15} adopted
the age of this neutron star from \citet{Tian_ea08}, who had suggested
association of the SNR with molecular gas connected with the HII region
G353.43$-$00.37 and calculated a radiative SNR age of 27~kyr. For that
age, the obtained ${\kB}T^\infty=153^{+4}_{-2}$ eV
would be too high to be explained by the
ordinary neutron star cooling theory. \citet{Klochkov_etal15} discussed
possible cooling scenarios including unusual baryon superfluidity to
explain this oddity.  Later, however, it was shown that, at $d=3.2$ kpc,
the SNR should be much younger, $t\sim2-6$~kyr \citep{Acero_15,CuiPS16,Maxted_ea18}, 
which can be consistent with less unusual cooling scenarios.

11. \emph{CXOU J181852.0$-$150213} is a CCO in the SNR G15.9+00.2. The
quoted luminosity and temperature correspond to the spectral fit with
the carbon atmosphere model for fixed $M=1.5\,\msun$ and $R=12$~km, with
unfixed $N_\mathrm{H}$ and $d$ \citep{Klochkov_ea16}. This fit gives
$d=10^{+9}_{-5}$ kpc, compatible with $d=8.5$\,--\,16 kpc derived by
\citet{Sasaki_ea18} from the SNR properties
(other fits give much larger $d$ and
therefore should be rejected). The age is estimated as
$t_*=(3.4\pm0.2)\,(d/10\mbox{ kpc})$ kyr  \citep{Sasaki_ea18}.

12. \emph{CXOU J185238.6+004020 (PSR J1852+0040)} is a CCO in the SNR
Kes 79 (G033.6+00.1). The age estimate is based the SNR  properties
\citep{Sun_ea04}. The luminosity and temperature are quoted from
\citet{Bogdanov14}, based on the \textsc{carbatm} model and assuming
$d=7.1$ kpc, $M=1.4\,\msun$ and $R=9$, 12, or 14 km. The effective
emitting area was allowed to vary and proved to be of the order of the
physical surface area (within a factor of $\sim0.8$\,--\,1.9). The 2BB
and \textsc{nsmax} fits give less plausible results (much smaller
emitting areas).

13. \emph{CXOU J232327.8+584842}, aka Cas A NS, is a CCO in the SNR Cas
A (G111.7$-$02.1), which was produced by supernova observed by Flamsteed
on August 16, 1680 (\citealt{Ashworth80}; this attribution is supported
by the age of the SNR, derived from observations by
\citealt{Fesen_ea06}). Historically, it is the first neutron star whose
spectrum has been successfully fitted by the carbon atmosphere model
\citep{HoHeinke09}. Its temperature and luminosity obtained using this
model are rapidly decreasing  over the time of its observations by the
\textit{Chandra}  X-ray observatory since 2000 \citep{HH10}. The
decrease of $T^\infty$ by  $\sim4-5$\% from 2000 to 2018 (see table~B1
in \citealt{Wijngaarden_19}) is at odds with theoretical cooling models.
It has been tentatively explained by neutrino emission in the Cooper
pairing processes (Sect.~\ref{sect:regulators}), assuming that the
internal temperature has passed the critical value for the onset of
neutron superfluidity a few decades ago
(\citealt{Shternin_etal11,PPLS11}; see also \citealt{Ho_ea15} and
references therein). However, a revision of the Cooper pairing-related
emission rate by \citet{Leinson10} makes the theoretical cooling rate
anyway incompatible with the observed one \citep{Leinson16,PC18}.
\citet{Posselt_ea13} suggested that the observed decline may be not
intrinsic to the neutron star, and put forward alternative hypotheses
for its explanation. According to \citet{PosseltPavlov18}, an assumption
that the column density can vary between the observations makes the
observed flux variation statistically insignificant. The analysis of
observations in Chandra ACIS Graded mode
\citep[e.g.,][]{HH10,Wijngaarden_19} gives noticeably higher $T^\infty$
and systematically quicker fading than those in Chandra ACIS subarray
mode \citep{Posselt_ea13,PosseltPavlov18}; this difference has not
been explained by the time of this writing. The ranges of $L^\infty$
and  $T^\infty$ in Table~\ref{tab:AgeL} accommodate different estimates
from the
above-cited works over the observation dates from 2000 to 2018.

\subsubsection{Moderately magnetized middle-aged pulsars}

14. \emph{PSR J0205+6449} is located in the SNR 3C 58 (G130.7+03.1),
probably a remnant of the historical supernova, observed starting from
August 6, 1181 (\citealt{Stephenson71}; the attribution was supported by
\citealt{Kothes13}). \citet{GreenGull82} measured a systemic velocity of
3C 58  about $-39$ km s$^{-1}$, which they translated to a distance of
2.6 kpc, in the Perseus spiral arm of the Milky Way. \citet{Roberts_93}
measured a systemic velocity of 3C 58 about $-36$ km s$^{-1}$, which
they translated to $d\sim3.2$ kpc, using another rotation curve of the
Milky Way. Having adopted this distance, \citet{Slane2004-PSRJ0205}
derived a limit of  $T^\infty < 1.02$ MK for blackbody emission from the
entire surface of the pulsar. They also showed that the data were
consistent with emission from \textsc{nsa} model for a canonical neutron
star with a similar temperature.

Using an advanced approach to the kinematic distance method developed by
\citet{FosterMacWilliams06}, \citet{Kothes13} derived the improved
distance estimate to 3C 58, $d=2.0\pm0.3$ kpc. This estimate is
confirmed by the trigonometric parallax measurements of the distance to
the HII region W3 \citep{Xu_06,Hachisuka_06}, which is located at
$d=2.0\pm0.1$ kpc a few degrees away from 3C 58 and has the same
systemic velocity. A spectral modelling of pulsar wind nebula around 3C
58 by \citet{TanakaTakahara13} is compatible with the distance of 2 kpc
for assumed ages of 1 kyr or 2.5 kyr.

We reanalysed \chan\ data using \textsc{tbabs} $\times$
(\textsc{nsmaxg}+PL) model with fixed $d=2$ kpc, $M = 1.4\msun$ and $R =
13$ km. We obtained  temperature from the entire NS surface  $T^\infty =
0.57^{+0.05}_{-0.07}$ MK (for the \textsc{nsmaxg} model 123100). Taking
the distance uncertainty into account, we obtain $L^\infty
= 1.9^{+1.5}_{-1.1}$ erg s$^{-1}$. 

15. \emph{PSR J0357+3205, named Morla.} \citet{DeLuca_ea13} put rigid
bounds on the distance between 0.2 and 0.9 kpc, but the likely
birthplace identified by \citet{Kirichenko_etal14} implies a narrower
uncertainty range $d=0.45\pm0.05$ kpc. The spectral analysis results in
Table~\ref{tab:AgeL} are also quoted from \citet{Kirichenko_etal14}, who
used a neutron star atmosphere model supplemented with power law
(NSA+PL). For the NSA component, the best fit ${\kB}T^\infty=31\pm1$ eV has
been obtained with the \textsc{nsmax} model
at fixed $M=1.4\,\msun$ and
$R=13$~km. This result for $T^\infty$ and the derived $L^\infty$ are
accepted in Table~\ref{tab:AgeL}. The errorbars embrace the result of
fitting with a non-fixed radius, which gives
${\kB}T^\infty=36^{+9}_{-6}$ eV and $\Reff=8^{+12}_{-5}$~km at the 90\%
confidence level. The luminosity uncertainties  in Table~\ref{tab:AgeL} ($1\sigma$),
derived from figures 2 and 3 of
\citet{Kirichenko_etal14}, also
accommodate the most likely values of other fits in table~2 of
\citet{Kirichenko_etal14}.

16. \emph{PSR J0538+2817} in the SNR Sim 147 (G180.0$-$01.7). The age is
determined by proper motion \citep{Kramer_03} and an analysis of the SNR
\citep{Ng_ea07}. 
The quoted luminosity and
temperature are mainly based on the \textsc{nsa} or \textsc{nsa}+PL
spectral fits by \citet{Ng_ea07} ($T^\infty=1.06\pm0.06$ MK,
$\Reff=11.2$~km). The uncertainties in the $L^\infty$ (but not in
$T^\infty$) in Table~\ref{tab:AgeL} also embrace the BB or BB+PL
results ($T^\infty=2.10\pm0.05$ MK, $\Reff\approx2.2$~km).

17. \emph{CXOU J061705.3+222127} resides in the SNR IC 443. Although
pulsations have not been detected, it is surrounded by a pulsar wind
nebula (PWN). \citet{Chevalier1999} has analysed
the SNR and obtained its age. Based on the PWN properties, \citet{Swartz2015-ic443}
derived  constraints on the period, $P\sim0.1-0.6$~s and the surface
magnetic field $B\sim4\times(10^{12}-10^{13})$~G. The fit for the
canonical neutron star with a hydrogen atmosphere \textsc{nsa} model
shows an effective temperature  ${\kB}T^{\infty} = 58.4^{+0.6}_{-0.4}$
eV and a  bolometric luminosity  $L^{\infty} = 
(2.6\pm0.1)\times10^{32}$ erg s$^{-1}$  \citep{Swartz2015-ic443} (the BB
fit gives $\Reff\approx1.6$ km, which is too small for radiation from
the entire surface but too large  for a hot spot).

18. \emph{PSR J0633+0632}. We adopt the results of
\citet{Danilenko_J0633}. The authors argue that the pulsar is hardly
associated with the previously suggested Monoceros Loop or Rosette
Nebula and point out the open stellar cluster Collinder 106 as another
possible birth site. For the thermal component of the spectrum, the
\textsc{nsmax} model of an orthogonal magnetic dipole looks most
realistic among several models tried by the authors. 
Magnetic field strengths $1.82\times10^{12}$~G or $10^{13}$~G at the
pole  yield similar results:
distance $d\sim0.9^{+1.1}_{-0.1}$ kpc, temperature
${\kB}T^\infty\sim53\pm8$~eV, a realistic radius of an equivalent
thermally emitting sphere $\Reff\sim10^{+17}_{-5}$~km, and luminosity
logarithm $\log L^\infty\mbox{(erg
s$^{-1}$)}\sim32.3\pm0.6$.  Here, the uncertainties are at the 90\%  confidence (in
Table~\ref{tab:AgeL} we have reduced them  to the $1\sigma$ level for
the sake of uniformity).  An alternative model of two hot spots using
the BB+PL fit  gives ${\kB}T^\infty=120\pm8$ eV,
$\Reff=0.8^{+0.5}_{-0.3}$~km, $\log L^\infty=31.4^{+0.3}_{-0.2}$,  and
$d=0.8^{+0.2}_{-0.1}$ kpc.

19. \emph{PSR J0633+1746, aka Geminga, a Musketeer.} The distance is
known from measured parallax. The best spectral fit by \citet{Mori_ea14}
with unfixed $N_\mathrm{H}$ consists of one BB and two PL components;
the resulting thermal luminosity range covers the one obtained by
\citet{DeLuca_ea05} with fixed $N_\mathrm{H}$ and 2BB+PL fit, after
scaling from $d=157$~pc (an older parallax measurement, used by
\citealt{DeLuca_ea05}) to the updated value of $d=250$~pc.

20. \emph{PSR B0656+14 (J0659+1414), a Musketeer} in the SNR Monogem Ring
(G201.1+08.3). The distance is known from measured parallax. The 2BB+PL
fit results by \citet{Arumugasamy_ea18} are similar to those
by \citet{DeLuca_ea05}. In Table~\ref{tab:AgeL} we
quote the results of the G2BBPL fit  of \citet{Arumugasamy_ea18}, that
is the 2BB+PL model with an added Gaussian absorption profile, which
gives similar (within 20\%) temperatures, but more realistic effective
radius of the cold component ($13^{+4}_{-3}$~km instead of $\sim20$~km).

21. \emph{PSR B0833$-$45 (J0835$-$4510)} resides in the SNR Vela (G263.9$-$3.3). 
The distance is known from measured parallax. Early estimates
of the true age ranged from 5 kyr to 50 kyr (see, e.g., \citealt{Stothers80}
for references), being highly model-dependent. 
Having analysed extended X-ray features in the SNR Vela,
\citet{AschenbachET95} derived $t_*=18\pm9$ kyr, 
which agrees with the independent estimate of $t_*\sim
18$ kyr by \citet*{JenkinsSW76}. Further analysis
\citep{Aschenbach02} showed that the SNR should be
at least $17-23$~kyr old. Therefore, we adopt the likely age interval to
be 17\,--\,27 kyr, which exceeds the canonical spindown
age $t_\mathrm{c}=\mathcal{P}/(2\dot{\mathcal{P}})=11.3$ kyr.  However,
the braking index of this pulsar is smaller than the canonical value
$n=3$. Before a glitch, $n=2.81\pm0.12$ \citep{Akbal_ea17}, which gives
the corrected spindown age (assuming negligibly small initial period)
$t_\mathrm{c}^*\equiv\mathcal{P}/[(n-1)\dot{\mathcal{P}}]=12.5\pm0.5$~kyr. 
When averaged
over a large timespan covering several glitches, the braking index
becomes $n=1.7\pm0.2$ \citep{EspinozaLS17}, which gives
$t_\mathrm{c}^*\sim(25-45)$~kyr. 
For the luminosity,
the 2BB spectral fit from table~6 of \citet*{ManzaliDLC07} yields
${\kB}T^\infty=93\pm3/186^{+5}_{-6}$ eV and
$\Reff=5.06^{+0.42}_{-0.28}/0.73^{+0.09}_{-0.07}$~km for the cold/hot
components at fixed $d=287$~pc, which gives
$L^\infty=3.1^{+0.5}_{-0.4}\times10^{32}$ erg s$^{-1}$; the hot
component can also be  fitted by a power law (PL) model. An NSA+PL fit at fixed
$M=1.4\,\msun$ and $R=10$~km gives $T^\infty=0.681\pm0.004$ MK and
$d=269^{+12}_{-14}$~pc, which corresponds to
$L^\infty=(2.61\pm0.06)\times10^{32}$ erg~s$^{-1}$. \citet{2018VELA}
reanalysed the pulsar spectrum and
obtained $T^\infty=0.700\pm0.005$ MK with the NSA+PL model
at fixed $d=287$~pc, $M=1.4\,\msun$, and $R=13$~km. The
employed \textsc{nsmax} model version assumes the dipole distribution of the magnetic field
over the surface and the ensuing distribution of temperature $\Ts$
\citep{HoPC08}. In
this case, the quoted $T^\infty$ is the effective temperature derived
from the total thermal luminosity, which implies 
$L^\infty=(4.24\pm0.12)\times10^{32}$ erg~s$^{-1}$.  Allowing for mass
and radius variations, \citet{2018VELA} arrived at the robust estimate
of the effective temperature $T^\infty=0.66^{+0.04}_{-0.01}$ MK with
$M=2.4^{+0.1}_{-1.4}\,\msun$ and $R=10.8^{+3.7}_{-1.3}$ km.

22. \emph{PSR B1055$-$52 (J1057$-$5226), a Musketeer.} The quoted
luminosity and temperature are based on the cold component of the 2BB+PL
fit by \citet{DeLuca_ea05}, which gives $T^\infty =
0.79\pm0.03$ MK, $\Reff=12.3^{+1.5}_{-0.7}$ km,
$L^\infty = 4.4\times10^{32}$ erg~s$^{-1}$ at fixed $d=750$~pc
(for the hot component, $T^\infty=1.79\pm0.06$~K,
$\Reff=0.46\pm0.06$ km,
$L^\infty = 1.6\times10^{31}$ erg~s$^{-1}$). Based
on an analysis of optical and ultraviolet observations,
\citet*{MignaniPK10} argue that the actual distance should be smaller
than the value $d=0.73\pm0.15$ based on the dispersion measure; they
give $d\sim200-500$ pc and scale luminosity to
$d=350$ pc. In Table~\ref{tab:AgeL} we show  $L^\infty$ with this scaling,
including the distance uncertainty. 

23. \emph{PSR J1357$-$6429} in the SNR HESS J1356$-$645 (G309.8$-$02.6).
The distance is evaluated by dispersion measure \citep{Zavlin07}. The
quoted luminosity and temperature correspond to the NSA+PL fit by
\citet{Chang_ea12}. An alternative BB+PL fit gives
$\Reff\sim2$~km, which is too small for the emission from the
entire surface \citep{Chang_ea12}.

24. \emph{PSR B1706$-$44 (J1709$-$4429)}, possibly in the SNR
G343.1$-$02.3. The distance is known from measured parallax. The
spectral analysis results are quoted for the best fit in
\citet{McGowan_etal04} (hydrogen NSA+PL model with fixed $R=12$~km);
uncertainties embrace other fits with the same $\chi_\nu^2=0.84$.

25. \emph{PSR J1740+1000.} The two distance values are given by two
different electron density distribution  models applied to the
dispersion measure. The spectral analysis results are quoted for the 2BB
fit \citep{Kargaltsev_ea12}. We present only results for the 2BB fit to
the ``rise and fall'' phase,  which are based on a larger energy
interval and yield plausible $\Reff\sim9^{+5}_{-3}$~km. The
best-fit value for the luminosity is taken from the 2BB model with fixed
${\kB}T^\infty=71$~eV. The uncertainties include phase
variations, 2BB fit with non-fixed $T^\infty$, and a contribution from
the hot component.

26. \emph{PSR J1741$-$2054 (Swift J174157.6$-$205411).}  The spectral
analysis results are quoted for the only acceptable BB+PL fit by
\citet{Karpova_etal14}, rescaled from $d=1$ kpc to 0.8 kpc, which yields
$\Reff=13.6^{+2.8}_{-2.4}$~km.

27. \emph{PSR B1822$-$09 (J1825$-$0935)}.
The X-ray spectrum 
is well described by a 2BB model with
$\kB T^{\infty}=83\pm4/187^{+26}_{-23}$ eV
and $\Reff=2.04^{+0.39}_{-0.37}/0.1^{+0.05}_{-0.03}$ km,
assuming a fixed distance $d=1$ kpc
\citep{Hermsen2017PSRB1822-09}. 
The hot component can be ascribed to the pulsed emission 
and the cool component to the unpulsed emission. 
By analogy with other pulsars,
we suggest that the bolometric flux,
which is mostly provided by the cool component,
is powered by cooling.

28. \emph{PSR B1823$-$13 (J1826$-$1334)} is associated with the
SNR G18.0$-$00.7. 
This glitching pulsar has a low time-averaged braking index
$n=2.2\pm0.6$ \citep{EspinozaLS17}, which gives the corrected
characteristic age $t_\mathrm{c}^*$ in the range 24\,--\,27 kyr.
The
spectral analysis results listed in Table~\ref{tab:AgeL} 
are adopted from \citet{Zhu_ea11}; they
represent the BB+PL fit of \citet*{PavlovKB08}. In this case,
$\Reff=5.1^{+0.4}_{-0.3}$~km.

29. \emph{PSR J1836+5925} (GRO J1837+59) was discovered in 1991
as a bright
$\gamma$-ray source \citep{GRO_J1837}. Later it was
identified as a neutron star and dubbed `Next Geminga' \citep{MirabalHalpern00}.
Its X-ray counterpart
was found in the \textit{ROSAT} data \citep{Reimer_01} and
observed with \chan\ and \xmm\ \citep{Arumugasamy2015PhD}. 
The distance estimate $d\approx250-750$ pc is based
on assumptions of $\gamma$-ray flux and beaming factor
with $\gamma$-ray pulse profile modelling \citep{Abdo_10_psrJ1836}. 
The BB+PL fit to the pulsar spectrum gives $\kB T^\infty= 63.6^{+4.7}_{-6.2}$
eV and $\Reff=1.55^{+0.61}_{-0.25} \dkpc$ km, whereas the fit
for the canonical neutron star model, using the
non-magnetic \textsc{nsa}+PL model, gives 
$\kB {T}_\mathrm{s}=20.7^{+6.5}_{-4.2}$ eV
(in Table~\ref{tab:AgeL} this result
is converted to the redshifted temperature and luminosity)
 and
$d=310^{+420}_{-170}$ pc at the 90\% confidence
\citep{Arumugasamy2015PhD}.

30. \emph{PSR B1951+32 (J1952+3252)} is associated with the SNR CTB 80
at $d\sim2$ kpc \citep{StromStappers00}. Proper motion of the pulsar
gives $t_*=64\pm18$ kyr \citep{Migliazzo_02}. 
The BB+PL fit to the pulsar spectrum gives 
$\kB T^\infty= 130\pm20$ eV 
and $\Reff=2.2^{+1.4}_{-0.8} (d/\mbox{2 kpc})$ km, whereas 
the 3$\sigma$ upper limit with fixed 
$\Reff=12$ km is 0.78 MK
 \citep{Li2005-psrb1951}. These values suggest 
 $L^\infty = 1.8^{+3.0}_{-1.1}\times10^{32}(d/\mbox{2 kpc})^2$ erg s$^{-1}$
and $L^\infty < 3.8\times10^{32}(d/\mbox{2 kpc})^2$ erg s$^{-1}$, respectively.

31. \emph{PSR J1957+5033} has been observed in gamma rays by
\textit{Fermi-LAT} \citep{FermiLAT2} and in X-rays by \textit{Chandra} 
\citep{Marelli_15} and by \textit{XMM-Newton}
\citep{Zyuzin_J1957hea,Zyuzin_J1957}. 
The `pseudo-distance' of this gamma pulsar, inferred from comparison
between flux and luminosity of gamma pulsars with known distance, equals
0.8 kpc \citep{Marelli_15}. The BB+PL fit yields
${\kB}T^\infty=56\pm7$~eV and $\Reff=(1.5-7.9)\dkpc$ km ($1\sigma$
confidence). The NSA+PL fit using the model of magnetized, partially
ionized
hydrogen atmosphere \textsc{nsmaxg} with fixed 
$M=1.4\,\msun$ and $R=13$~km for the thermal component results in
temperature close to the lowest value $\log \Ts\mbox{(K)}=5.5$ available for these models
\citep{Zyuzin_J1957hea}. 
\citet{Zyuzin_J1957} have computed new atmosphere models,
which include lower temperatures, assuming dipole magnetic fields.
The polar magnetic field strengths $\sim(1-3)\times10^{12}$~G
are chosen to be consistent with timing
for this pulsar. The atmosphere models
were built for several sets of neutron star parameters. Depending on
these parameters
and on the inclination of the magnetic dipole axis to the line of sight,
the estimates range from ${\kB}T^\infty=14\pm3$ eV to $21\pm4$ eV
at distances $\sim50-300$ pc
\citep{Zyuzin_J1957}.

32. \emph{PSR J2021+3651} powers the Dragonfly Nebula (PWN 75.2+0.1).
The X-ray spectra of the pulsar and the nebula were resolved 
in the \textit{Chandra} observations by \citet*{VanEttenRN08}.
\citet{Kirichenko_15} conducted deep optical observations of this
complex with the
\textit{Gran Telescopio Canarias} and reanalysed the archival \textit{Chandra} X-ray
data. They constructed the extinction-distance relation for the
direction toward the pulsar and constrained the distance 
$d=1.8^{+1.7}_{-1.4}$~kpc at the 90\%
confidence. The BB+PL fit to the pulsar spectrum gives $\kB T=155\pm14$
eV and $\Reff=1.3^{+1.5}_{-1.0}$ km, whereas the NSA+PL fit, using the
\textsc{nsmax} model, gives $\kB T=63^{+9}_{-8}$ eV and
$\Reff=12.0^{+19.5}_{-9.6}$ km at the 90\% confidence. In Table~\ref{tab:AgeL}
we reduce the
errors to the $1\sigma$ confidence level for uniformity with other data
in the tables. These fits correspond to 
$L^\infty \sim 10^{31} - 10^{33}$ erg s$^{-1}$,
respectively. The main source of uncertainty is the poorly known
distance.  For fixed $R=13$ km and
$M=1.4\,\msun$, the NSA+PL fit
gives $L^\infty = 5^{+3}_{-2}\times10^{32}$ erg s$^{-1}$.

33. \emph{PSR B2334+61  (J2337+6151)} in the SNR G114.3+0.3. The 
NSA fit by \citet{McGowan_ea06} for the canonical
neutron star model with $B=10^{13}$~G
gives $d=1.1\pm0.6$ kpc and non-redshifted temperature
${T}_\mathrm{s}=0.58^{+0.13}_{-0.25}$ MK at 90\% confidence, which
 converts into $T^\infty = 38^{+6}_{-9}$ eV and
$L^\infty=(4.7\pm3.5)\times10^{31}$ erg s$^{-1}$ ($1\sigma$). 
The authors prefer a
fit with $R=13$~km, because it yields a distance compatible with
$d=3.1^{+0.2}_{-1.0}$ kpc derived from the pulsar dispersion measure
using the NE2001 model of the Galactic electron distribution
\citep{CordesLazio02,CordesLazio03}. It gives the non-redshifted
${T}_\mathrm{s}=0.65^{+0.13}_{-0.34}$ MK at 90\% confidence, which
at the $1\sigma$ converts into $T^\infty = 46^{+6}_{-16}$ eV and
$L^\infty=(1.5^{+1.0}_{-1.2}\times10^{32}$ erg s$^{-1}$.  However, the
NE2001 model 
incorporates a void in the direction of G114.3+0.3, likely to
accommodate for the larger distance to the SNR predicted previously by
\citet{ReichBraunsfurth81}. The modern model of the Galactic
distribution of free electrons \citep{YaoMW17} yields the distance to
the pulsar of 2.08 kpc. Moreover, \citet{Yar-Uyaniker2004-psrb2334}
suggested the distance to the SNR of about 700 pc from HI data analysis,
which is compatible with the result of the canonical neutron star NSA
fit mentioned above. We should note that the BB fit gives much higher
temperature $T^\infty=1.62\pm0.23$ MK and small effective radius
$\Reff\sim0.5\dkpc$ km, suggesting a possible alternative
interpretation of the thermal radiation as produced by hot spots.

\subsubsection{High-$B$ pulsars and XINSs}

34. \emph{PSR J0726$-$2612}.  The reported luminosities and
temperatures correspond to the best G2BB fit of \citet{Rigoselli_J0726}. The
dispersion measure implies $d = 2.9$ kpc assuming the Galactic
electron distribution, but it may be an overestimate, as discussed by
\citet{Rigoselli_J0726}. \citet*{SpeagleKVK11}  suggested that PSR
J0726$-$2612 could be associated with the Gould belt and hence
$d\lesssim1$ kpc. The spectral fitting has been performed for fixed
$d=1$ kpc. Then the inferred effective radius for the cooler component
in the G2BB fit is $\Reff=10.4^{+10.8}_{-2.8}$ km.

35. \emph{PSR J1119$-$6127} resides
in the SNR G292.2$-$0.5. The NSA+PL fit of the
phase-averaged spectrum with fixed $R=13$~km yields
${\kB}T^\infty=80^{+30}_{-20}$ eV and $d=2.4^{+5.6}_{-1.8}$ kpc, while the
BB+PL fit at fixed $d=8.4$ kpc gives ${\kB}T^\infty=210\pm40$ eV and
$\Reff=3^{+4}_{-1}$ km \citep{Ng_ea12}. Interestingly, for
braking index $n=2.684$ of this pulsar, its hosting SNR age of
4.2\,--\,7.1 kyr, evaluated
at $d=8.4$ kpc, significantly exceeds the corrected characteristic age 
$t_\mathrm{c}^*=1.9$~kyr, which is usually considered as an upper limit
to the pulsar's age \citep*{KumarSHG12}. We note that adopting the
best-fit distance $d=2.4$ kpc from the spectral analysis would reduce
the estimated age to 1.2\,--\,2.1 kyr, in agreement with the
characteristic age. However, as follows from the analysis  by
\citet{Caswell_ea04}, such a short distance should imply an unusually
high interstellar absorption in the pulsar direction. It is more likely that
the mean braking index may be smaller after averaging over a long time
covering many glitches, which is usual for the glitching pulsars like this
one (see \citealt{EspinozaLS17}); then a lower $n$ implies a
larger $t_\mathrm{c}^*$.

36. \emph{PSR B1509$-$58 (J1513$-$5908)} has a measured braking index
$n=2.832$ \citep{LivingstoneKaspi11}, which gives a corrected characteristic age
$t_\mathrm{c}^*=1.7$~kyr. The temperature is quoted from Table~4
of \citet{Hu_ea17}. It is obtained with BB+PL fitting for fixed $d=5.2$
kpc and $\Reff=13$ km. The luminosity in our Table~\ref{tab:AgeL}
corresponds to these data. However, $\Reff$ is poorly constrained by the
observations, and fig.~10 of \citet{Hu_ea17} shows a larger luminosity
interval, $1.0\times10^{33}-1.5\times10^{34}$ erg~s$^{-1}$.

37. \emph{PSR J1718$-$3718.} The luminosity and temperature estimates
are quoted from \citet{Zhu_ea11}. The BB model preferred by the authors
leads to ${\kB}T^\infty=186^{+19}_{-18}$ eV, $\Reff=1.8^{+1.7}_{-0.6}(d/\mbox{4.5 kpc})$ km
and $L^\infty=4^{+5}_{-2}\times10^{32}(d/\mbox{4.5 kpc})^2$ erg s$^{-1}$. 
 On the other hand, the NSA model with the canonical neutron star parameters,
assuming $d=4.5$\,--\,10 kpc, leads to $\kB{T}_\mathrm{s} = 75$\,--\,97 eV, which
gives $\kB T^\infty=57$\,--\,74 eV and 
$L^\infty=(2.4-6.7)\times10^{32}$ erg s$^{-1}$.

38. \emph{PSR J1819$-$1458} is the only `rapid radio transient'
registered in X-rays (see \citealt{Gencali_Ertan19} and references
therein). The 2GBB fit by \citet{Miller_ea13} (the only one with
$\chi_\nu^2<1.1$ at $d=3.6$ kpc) gives ${\kB}T^\infty=138.2\pm0.9$ eV
and $\Reff=8^{+5}_{-4}$ km. The inferred value of $L^\infty$  in our
Table~\ref{tab:AgeL} is consistent with the unabsorbed bolometric flux
for this model. The errors accommodate those alternative spectral fits
(BB and GBB) in Table~2 of \citet{Miller_ea13} that have plausible
$\Reff<20$ km.

39. \emph{RX J0420.0$-$5022}. The reported results are based on the
best fit by \citet{Haberl_ea04} (GBB model for the canonical neutron
star) and the results listed by \citet{KvK09}.

40. \emph{RX J0720.4$-$3125}. The spin period of this object is the
longest among all currently known XINSs (\citealp{Hambaryan_17}; the
period was
previously thought to be twice shorter because of comparable pulses from
two antipodal spots). This XINS shows significant  variability of its
X-ray spectrum \citep{Hohle_ea12,Hohle_ea12b}.  Its distance is known
from measured parallax. An analysis of \textit{Chandra} co-added spectra
with the GBB model \citep{Hohle_ea12} yields $L^\infty$ and
$T^\infty$ quoted in Table~\ref{tab:AgeL} with
$\Reff=4.5^{+1.3}_{-1.1}$ km at $d=0.3$ kpc. 
\citet{Hambaryan_17} performed a phase-resolved spectral analysis  with
more physical models of condensed surface and magnetized atmosphere, 
taking a non-uniform temperature distribution into account. They found
phase-dependent best-fit temperature values at magnetic poles
in the range ${\kB}T^\infty\approx 98-115$ eV,

41. \emph{RX J0806.4$-$4123}. The BB and GBB fits by \citet{Haberl_ea04}
give ${\kB}T^\infty=104\pm4$ eV 
and $92\pm4$ eV, respectively. The luminosities are derived assuming
$\Reff=1.3$ km as given by \citet{KvK09} for the latter fit.
They agree with the luminosities in \citet{Vigano_13}.

42. \emph{RX J1308.6+2127 (RBS1223)}. The distance $d\sim380$ pc has
been obtained from the spectral analysis together with the temperature
and luminosity by \citet{Hambaryan_ea11}. Of three different age estimates
suggested by \citet{Motch_ea09}, only the one quoted in
Table~\ref{tab:AgeL} is compatible with this distance. For $d=380$ pc,
the spectral fit implies a large non-redshifted radius $R=16\pm1$ km,
which may suggest that the actual distance is closer to the estimate by
\citet{Motch_ea09} for the possible neutron-star birthplace,
$d=260\pm50$ pc. Alternatively this may imply
a higher mass, which would yield a large $\Reff$.
The effective temperature for this radius is 
$7\times10^5$~K in the local reference frame.
The corresponding luminosity at infinity is
$L^\infty=(3.3\pm0.5)\times10^{32}$ erg s$^{-1}$. 
 The model temperature distribution over the surface has 
maximum at ${\kB}\Ts=105^{+2}_{-4}$~eV 
 (the minimum is much lower).
 An alternative BB fit for the canonical
neutron star model gives 
${\kB}\Ts=100$ eV and dilution factor 0.34,
which corresponds to $L^\infty\sim2.6\times10^{32}$ erg s$^{-1}$.

43. \emph{RX J1605.3+3249 (RBS1556)}. 
The age estimate is based on Table~4 of 
\citet{Tetzlaff_ea12}. 
\citet{Posselt_ea07}, using different models 
of the hydrogen column density,
derived distances of 390 pc and 325 pc.
On the other hand, \citet{Motch_05}
link the source with the Sco OB2
association within the Gould Belt, at a mean distance of
120\,--\,140 pc. \citet{Tetzlaff_ea12} argue that the neutron star was
probably born in the Octans association from a supernova
at $d=140^{+6}_{-19}$ pc. They adopt the current distance of 300\,--\,400 pc
from \citet{Posselt_ea07}, which requires rather large (though not impossible)
space velocity $\sim550$ km s$^{-1}$.
\citet{Pires_ea19} have performed a timing and spectral analysis of \xmm\
observations. The best multi-epoch fit G2BB at
fixed $d=300$ pc for the cooler component gives
${\kB}T^\infty=60.9^{+1.7}_{-1.5}$~eV and
$\Reff=16.2^{+0.6}_{-1.1}$~km, in which case
$L^\infty=(4.7\pm0.5)\times10^{32}$ erg s$^{-1}$ (which we adopt as an upper 
bound); the hotter
component with ${\kB}T^\infty\approx117$~eV and $\Reff\approx1.34$~km
adds about 10\% to the total energy flux. 
Multi-epoch fits by different NSA models,
modified by a broad Gaussian absorption line at energy $385\pm10$ eV,
give $d\sim110$\,--\,130 pc and $\log \Ts\mbox{(K)}\approx5.6$\,--\,5.8,
corresponding to $\log L^\infty(\mbox{erg s$^{-1}$})\sim31.4\pm0.6$
(providing the lower bounds in Table~\ref{tab:AgeL}).
Note that
\citet{Pires_ea19} have disproved a previously reported spin periodicity
of this X-ray source. \citet{Malacaria_19}
performed a joint analysis of the \textit{NICER} and
\textit{XMM-Newton} data. 
These authors found that $d\sim350$ pc is hard to accommodate with their data,
while a possible distance $\sim100$\,--\,200 pc is consistent with the data.
The G2BB model gives $T^\infty=63^{+7}_{-6}/119^{+6}_{-4}$ eV and
$L^\infty\sim(3-5)\times10^{32}$ erg s$^{-1}$.
Atmosphere models \textsc{nsa} and \textsc{nsmaxg}, modified by 
an absorption line at $\sim450$ eV, yield similar
effective temperatures ${\kB}{T}_\mathrm{s}=47.0\pm0.5$ eV (for fixed $M=1.4\,\msun$ and $R=10$ km,
which requires $d=92\pm5$ pc)
and ${\kB}{T}_\mathrm{s}=46.2^{+1.7}_{-2.3}$ eV (at fixed $d=100$ pc,
which gives
$M=2.04^{+0.19}_{-0.49}\msun$, $R=15.6^{+0.6}_{-0.8}$ km; all errors are
at 90\% confidence).
These results lead to $L^\infty=(3.7\pm0.2)\times10^{31}$ erg s$^{-1}$
(\textsc{nsa}) or
$L^\infty=(8.8\pm1.0)\times10^{31}$ erg s$^{-1}$
(\textsc{nsmaxg}), within the bounds provided by the analysis of \citet{Pires_ea19}.

44. \emph{RX J1856.5$-$3754}, aka the Walter star, is the first
discovered neutron star with purely thermal spectrum
\citep*{WalterWN96}. Its likely birthplace is Upper Scorpius OB
association, which gives $t_*=420\pm80$ kyr
\citep{Mignani_ea13}. The distance is known from parallax measurements.
The spectral analysis performed by \citet{Ho_etal07} with the model of a
thin partially ionized hydrogen atmosphere with magnetic field
$B\sim(3-4)\times10^{12}$~G, being scaled to the updated 
distance $d=123^{+11}_{-15}$ pc \citep{Walter_ea10},
leads to $\kB T^\infty=37.4\pm0.3$ eV, 
$R=12.1^{1.3}_{-1.6}$ km, $M=1.48^{+0.16}_{-0.19}\,\msun$ and
$L^\infty=5.8\pm0.2$ erg s$^{-1}$ \citep{Potekhin14}. Alternative 2BB,
2BB+PL, and 3BB fitting models have been presented by
\citet{Sartore_ea12} and \citet{Yoneyama_ea17}. They yield 
${\kB}T^\infty=39^{+5}_{-3}/62.4^{+0.6}_{-0.4}$ eV and
$\Reff\sim12/4.7$ km for two BB components, which give similar 
luminosities at infinity, $\sim4\times10^{31}$
erg~s$^{-1}$ each.

45. \emph{RX J2143.0+0654 (RBS1774, 1RXS J214303.7+065419)}. The quoted
luminosities and temperatures are based on the spectral analysis by
\citet{Schwope_ea09}, who supplemented the joint \chan\ and \xmm\
observations with deep optical observations. The lower luminosity bound
$L^\infty=6.3\times10^{31}$ erg s$^{-1}$ corresponds to the colder
component (${\kB}T^\infty=40$ eV, $\Reff=13.8$ km) at the lowest
possible distance $d=250$ pc. The hot circular spot with radius of 1.6
km and ${\kB}T^\infty=104$ eV adds approximately $10^{31}$ erg s$^{-1}$,
which should be doubled  for two antipodal hot spots. The colder
component rescaled to fiducial distance $d=410$ pc corresponds to
$L^\infty\approx1.7\times10^{32}$ erg s$^{-1}$, which is adopted as the
upper limit in Table~\ref{tab:AgeL}.  By order of magnitude, these
estimates agree with \citet{Zampieri_ea01}, who found
$L^\infty\sim10^{32}\,(d/300\mbox{~pc})^2$ erg s$^{-1}$ On the other
hand, if the optical flux measured by \citet{Schwope_ea09} has a
non-thermal origin, then one can rely on the analysis of this source by
\citet{Cropper_ea07}, based on BB or GBB models. At fixed $d=300$ pc it gives
${\kB}T^\infty\approx101-104$ eV, $\Reff\approx2$ km, and
$L^\infty\sim(5-6)\times 10^{31}$ erg s$^{-1}$ (we adopt it as a lower bound).
Alternative fits for the canonical neutron star model and magnetized NSA
give smaller temperatures ${\kB}T^\infty\sim24-40$ eV and luminosities
$L^\infty\approx(0.7-6)\times10^{31}$ erg s$^{-1}$, but show a poor
statistical significance $\chi_\nu^2\sim2$ \citep{Cropper_ea07}.

\subsubsection{Upper limits on cooling-powered thermal emission}

46. \emph{PSR J0007+7303} is a radio-quiet pulsar associated with the
SNR CTA~1 (G119.5+10.2). It was observed in X-rays with \chan,  \xmm\
and \textit{Suzaku}
\citep{Halpern_etal04,Caraveo2010-psrj0007,Lin2010-psrj0007,Lin_12} and
in gamma-rays with \textit{Fermi-LAT} \citep{Abdo_12}. The distance of
$1.4\pm0.3$ kpc is estimated from the velocity of an H I shell
associated with the SNR CTA~1 \citep{Pineault_93}. The  age of 9.2 kyr
is derived from  modelling the dynamics and spectra of the pulsar
wind nebula and the SNR CTA~1 using estimates of the molecular mass in
the vicinity of the complex \citep{MartinTP16}. The analysis of the pulsar X-ray spectrum 
has been performed by \citet{Caraveo2010-psrj0007}.
The BB+PL fit gives
${\kB}T^\infty = 102^{+32}_{-18}$ eV, $\Reff=0.64^{+0.88}_{-0.20}d_{1.4}$ km
and thermal luminosity 
$L^\infty = (3.6\pm1.4)\times10^{30}d_{1.4}^2$ erg s$^{-1}$,
where $d_{1.4}=d/\mbox{(1.4 kpc)}$.
The \textsc{nsa}+PL fit gives ${\kB}T^\infty=54^{+25}_{-16}$ eV, 
$\Reff=4.9^{+1.8}_{-4.7}d_{1.4}$ km and 
$L^\infty = 3.9^{+2.6}_{-1.3}\times10^{30}d_{1.4}^2$ erg s$^{-1}$.
A power law without a thermal component also provides
a good fit ($\chi_\nu^2=0.74$), therefore the above estimates
of thermal luminosity can only provide an upper limit.

47. \emph{PSR B0531+21} is located in the Crab Nebula, a remnant of the historical
supernova, observed starting from July 4, 1054
\citep[e.g.,][]{StephensonGreen03}. \citet{Trimble73} estimated a range of distances 
between 1.4 and 2.7 kpc based on a variety of lines of evidence. 
The $L^\infty$ and $T^\infty$ limits in Table~\ref{tab:AgeL}
are determined from Fig.~5 
of \citet{Weisskopf_ea11} at the
$3\sigma$ confidence level, assuming $M=1.4\,\msun$, $R=12$ km,
and $d=2$ kpc.

48. \emph{PSR B1727$-$47 (J1731$-$4744)} is located in the SNR RCW 114
(G343.0$-$06.0). \citet{Shternin_ea19} measured the proper motion of the pulsar
and determined its likely birthplace, distance and age.
The spectral analysis has been performed by \citet{Zyuzin_B1727} using the 
\textsc{nsmaxg}+PL model
with fixed  $d=0.75$ kpc and $R=12.5$~km.

49. \emph{PSR J2043+2740}\label{J2043} is located near the edge of the
Cygnus Loop (SNR G074.0-08.6), whose age is $\sim10$--20 kyr at distance $735\pm25$ pc
\citep{Fesen_18}. However, the association with this SNR is very
uncertain. Therefore, the distance based on the dispersion measure is
usually adopted: $d=1.8$ kpc in early works
\citep{Becker_04,ZavlinPavlov04,Zavlin09} or $d=1.48$ kpc in more recent
papers \citep{Testa_18}. This pulsar has also been observed in gamma
rays \citep{FermiLAT2} and in optical bands 
\citep{Beronya_15,Testa_18}; its multi-wavelength optical-to-gamma
spectrum has been discussed by \citet{Testa_18}. There was only one
12-ks observation of this pulsar in X-rays by \textit{XMM-Newton} in
2002, which gathered about one hundred counts. \citet{ZavlinPavlov04}
and \citet{Zavlin09} argued that this emission should be mostly thermal,
since a PL fit gave implausibly large photon index $\sim5$. For a fixed
$d=1.8$ kpc, these authors obtained effective temperatures ranging from
0.5 to 0.9 MK and emitting radii from $\sim9$ down to $\sim2$ km, with
$L^\infty\sim(2-4)\times10^{31}$ erg s$^{-1}$, depending on a fit model.
For a hot spot with radius $0.47$ km, inferred from the
magnetic dipole model \citep[e.g.,][]{ManchesterTaylor}, they obtained
$L^\infty\sim10^{30}$ erg s$^{-1}$.
However, this last estimate would imply an implausibly small distance
$d\lesssim0.4$ kpc. On the other hand, an analysis of the same data by
\citet{Becker_04} showed a poor statistics for the BB or 2BB fits, but
gave an acceptable PL fit with power index of $3.1^{+1.1}_{-0.6}$.
Because of all these uncertainties, we treat the maximum temperature and
luminosity estimates derived by \citet{ZavlinPavlov04} as upper limits.
A longer observation of this object would be desirable to shed light on
the origin of its X-ray emission.

50. \emph{PSR B2224+65 (J2225+6535)} is associated with the Guitar 
bow-shock H$\alpha$ nebula. 
Its radio parallax is $1.20^{+0.17}_{-0.20}$ mas \citep{Deller_19}.
Only non-thermal X-ray flux has been registered from this pulsar.
\citet{HuiBecker07} estimated its X-ray luminosity as
$L_X\sim(1-2)\,\dkpc^2 \times10^{30}$ erg s$^{-1}$. It was
not clear, whether the bulk of these observed X-rays originated from the
pulsar magnetosphere or from the pulsar wind nebula \citep{Hui_ea12}.
The
analysed data only include photon energies above 0.7 keV, which leaves a
bolometric correction very uncertain. \citet{HuiBecker07} also estimated
the upper bound on the temperature of a hot spot $T^\infty<1.3$~MK by
adding a BB component to the best-fit PL model and
assuming a polar cap of radius 175~m, derived from the standard dipole
model (e.g., \citealt{ManchesterTaylor}). We adopt this constraint as a
conservative upper limit to the temperature. A more restrictive estimate
$T^\infty<0.61$ MK ($3\sigma$) was derived by  \citet{HuiBecker07}
assuming that the thermal flux is emitted from the whole surface of
a canonical neutron star. It corresponds to bolometric flux
$L^\infty < 1.7\times10^{32}$ erg s$^{-1}$, which we take as a
conservative upper limit to the thermal luminosity.

\subsubsection{Hot spots on the surfaces of old rotation-powered pulsars}

51. \emph{PSR B0114+58 (J0117+5914)}. The distance $d=1.77\pm0.53$ kpc
has been inferred from the dispersion measure by
\citet{RigoselliMereghetti18}.  The authors obtained the quoted
$T^\infty$ and $L^\infty$ by an analysis of archival \textit{XMM-Newton}
observations with the BB model. They estimated the effective radius of a
plane hot spot on the stellar surface to be $450^{+110}_{-90}$~m; in
Table~\ref{tab:AgeL} we list half of this number for the radius of an
equivalent sphere $\Reff$.

52. \emph{PSR B0943+10 (J0946+0951)}. We mainly rely on the spectral
analysis by \citet{Rigoselli_B0943}. Namely, we have selected a fit
model of the partially ionized hydrogen atmosphere with
$B\approx2\times10^{12}$~G in the so called B-mode, where the X-ray flux
has minimum. This fit yields ${\kB}T^\infty=82^{+3}_{-9}$ eV and
$\Reff=170^{+45}_{-25}$~m.  The condensed surface models give
${\kB}T^\infty\sim200-220$ eV and $\Reff\sim40-60$~m. The
thermal component of the BB+PL fit  yields ${\kB}T^\infty=210\pm20$ eV
and $\Reff=41^{+10}_{-9}$~m. The corresponding luminosity range
is shown in Table~\ref{tab:AgeL}; it agrees with the  range of the
unabsorbed X-ray flux given in that reference.

53. \emph{PSR B1133+16 (J1136+1551)}. The distance is known from
measured parallax. Over 2/3 of this pulsar's luminosity in the
0.3\,--\,2 keV range is non-thermal; the radius of equivalent emitting
sphere is $\Reff=17^{+7}_{-5}$~m \citep{Szary_ea17}. The redshifted
luminosity and temperature in Table~\ref{tab:AgeL} are derived from
the non-redshifted values in table~6 of that paper.

54. \emph{PSR J1154$-$6250}. The listed distances correspond to two
models of the Galactic electron density distribution; the pulsar's
projection on the Cru OB1 association is most likely a chance
coincidence \citep{Igoshev_ea18}. The quoted bolometric luminosity is
obtained from the unabsorbed flux $(7.5\pm2.2)\times10^{-15}$ erg
cm$^{-2}$ s$^{-1}$ at $d=1.36$ kpc (table 3 of the cited reference). It
assumes thermal interpretation. However, it may turn out to be only an
upper limit, since the PL model fits the spectrum equally well
\citep{Igoshev_ea18}.

55. \emph{PSR B1929+10 (J1932+1059)}. The distance is known from
measured parallax. The quoted thermal luminosity from 
\citet*{MisanovicPG08} is scaled to the updated distance $d=310$ pc
\citep{Verbiest_ea12} from $d=361$~pc adopted by the authors; the scaled
radius of an equivalent emitting sphere$\Reff = 28^{+5}_{-4}$~m
is obtained for the BB+PL spectral model. It may turn out to be only an
upper limit, since the PL model yields an acceptable fit with only
slightly larger $\chi_\nu^2$.

\subsection{Excluded objects}

Here we do not consider the soft gamma repeaters and anomalous X-ray
pulsars (SGR/AXPs), luminous neutron stars which reveal powerful bursts
Probably they are magnetars, neutron stars with superstrong magnetic
fields $B\sim10^{14}-10^{16}$~G, which power their bursting activity
(see, e.g., \citealt*{2015Mereghetti,2017KasB}, for recent reviews).
Their persistent radiation, albeit thermal-like in the soft X-ray band,
can hardly be related to passive cooling \citep[see][]{Vigano_13,PC18}.
It is thought to arise from a complicated blending of surface thermal
emission distorted by the presence of a highly magnetized atmosphere,
then Comptonized by currents in the magnetosphere, which can further
result in surface heating via return currents (e.g., \citealt{2017KasB},
and references therein). The surface may be also heated by energy
release in the crust, driven, for example, by magneto-thermal
\citep[e.g.,][]{Vigano_13} or magneto-thermoplastic evolution
\citep*{LiLB16}.

As mentioned in Sect.~\ref{sect:gen}, the millisecond pulsars  cannot be
used for direct testing the cooling theory. They have been recycled
during  accretion from a binary companion (e.g., \citealt{Bisno06} and
references therein), and their ages (typically of the order of
gigayears) greatly exceed the passive cooling timescale. Therefore,
thermal radiation registered from the millisecond pulsars originates
from late-stage heating, either the heating of hot spots by fast
particles from the magnetosphere or the internal heating due to slow
non-equilibrium processes (\citealt{GonzalezReisenegger10} and
references therein). The heating hypothesis was supported by the
detection of the far-UV  part of the thermal emission from the bulk of
the surface of the closest ($d=156.3$ pc) millisecond pulsar
J0437$-$4715 \citep{2004kargaltsev,2012durant} with $t_\mathrm{c}=6.64$
Gyr and possibly also from the surface of the millisecond pulsar
J2124$-$3358 \citep{Rangelov_17} with Lutz--Kelker-bias-corrected
distance $d=300^{+70}_{-50}$ pc \citep{Verbiest_ea12} and
proper-motion-corrected $t_\mathrm{c}=10.7$ Gyr \citep{ATNF}. Recently,
modelling the cool thermal component of the UV\,--\,X-ray spectrum of 
PSR J0437$-$4715 has resulted in the estimates $R=13^{+0.9}_{-0.8}$ km
and $T^\infty=(2.3\pm0.1)\times10^5$~K, which correspond to
$L^\infty=(5.3\pm1.0)\times10^{30}$ erg s$^{-1}$
\citep*{GonzalezCaniulef_GR19}. Upper limits to thermal fluxes from some
millisecond pulsars can be found in \citet*{GJPR15}. A recent study of 
the millisecond pulsar J0952$-$0607 \citep{HoHC19} provides an upper
bound on its luminosity $L^\infty\lesssim10^{31}$ erg s$^{-1}$. Thermal
emission has also been identified from hot ($kT^\infty=260^{+30}_{-20}$
eV) polar cap of the very old ($t_\mathrm{c}=16.5$ Gyr) millisecond
pulsar J1909$-$3744 with well known mass and distance, which gives
$L^\infty\approx1.5\times10^{30}$ erg s$^{-1}$ \citep{Webb_19}.

We have discarded PSR B0355+54, PSR B1916+14, and PSR J1734-3333 from
the catalogue of \citet{Zhu_ea11}, because their effective temperatures
and thermal luminosities (probably of hot spots, as suggested by small
effective radii) appear to be poorly constrained. We have also discarded
PSR B0950+08 and PSR B0823+26 from the same catalogue. In the case of
PSR B0823+26, recent observations and analysis \citep{Hermsen_ea18}
reveal a small effective emitting area ($\Reff<100$ m) for a
thermal-like component of the spectrum; moreover, this component is only
observed in the `bright mode' of the pulsar (in the `null mode' the
X-ray flux is below detection threshold), which excludes its
interpretation in terms of passive cooling. In the case of PSR B0950+08,
its effective surface temperature   $\sim(1-3)\times10^5$~K and
bolometric thermal luminosity $L^\infty=8^{+7}_{-4}\times10^{29}$ erg
s$^{-1}$,  obtained recently by \citet{Pavlov_ea17}, should be caused by
reheating at its characteristic age $t_\mathrm{c}=17.5$ Myr (as for the
above-mentioned millisecond pulsars), whereas treating $t_\mathrm{c}$ as
only an upper limit to the true age makes this constraint too loose to
be useful. 

Recently, \citet{Guillot_19} have obtained an upper bound
$\Ts<4.2\times10^4$~K for slowly rotating ($\mathcal{P}=8.51$~s) old
(proper-motion-corrected $t_\mathrm{c}=333$ Myr) PSR J2144$-$3933, which
makes it the coldest known neutron star and indicates that the
integrated power of reheating processes in slowly rotating neutron stars
may be below $10^{28}$ erg s$^{-1}$. We have not included
this result in the table, because it does not constrain
the theory of passive cooling.

The neutron star candidate 1WGA J1952.2+2925 in PWN DA 495  near
the center of SNR G065.7+01.2 has a pure thermal spectrum
\citep{Karpova2015MNRAS-da495}, but its age, distance and temperature
are very uncertain.
Pulsars J0554+3107 and J1105-6037 show thermal emission in
the X-ray spectra \citep{Zyuzin2018MNRAS-6psrs}, however, due to small
count statistics it is hard to estimate  their thermal luminosities. The
X-ray source RX J0002.9+6246, which was listed in several cooling
neutron star collections starting from \citet{Page_04}, has been also
discarded, because it turned out to be an ordinary star
\citep{Esposito_08}.

\begin{figure}
  \centering
  \includegraphics[width=.48\textwidth]{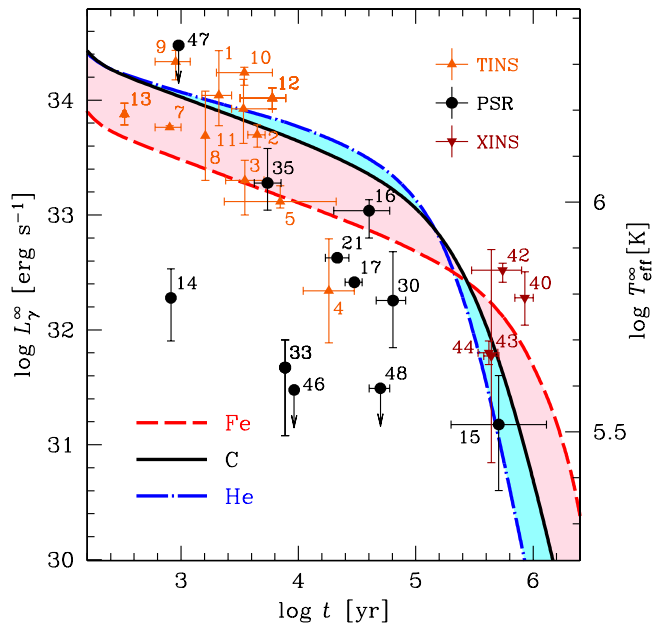}
  \caption{Cooling curves of neutron stars with $M=1.4\,\msun$,
  $R=12.6$~km and  different 
  heat blanketing envelopes: purely ground-state (dashed curve) and
  replaced with carbon (solid curve) or helium (dot-dashed curve).
  The left vertical axis is the thermal photon luminosity, the right axis is 
  the effective temperature,
  as seen by a distant observer. The curves are
  compared with the data on INSs with estimated ages and luminosities from
  Table~\ref{tab:AgeL}. The data are plotted as
  indicated in the legend (color online), 
  for different neutron star classes: thermally emitting INSs 
  (TINSs, including CCOs), pulsars (PSRs, including high-B
  pulsars), and XINSs. The error bars show uncertainties (typically $1\sigma$)
  and the arrows correspond to upper limits (at $3\sigma$ confidence).
  }
  \label{fig:coolsfx}
\end{figure}
\begin{figure*}
\includegraphics[width=.48\textwidth]{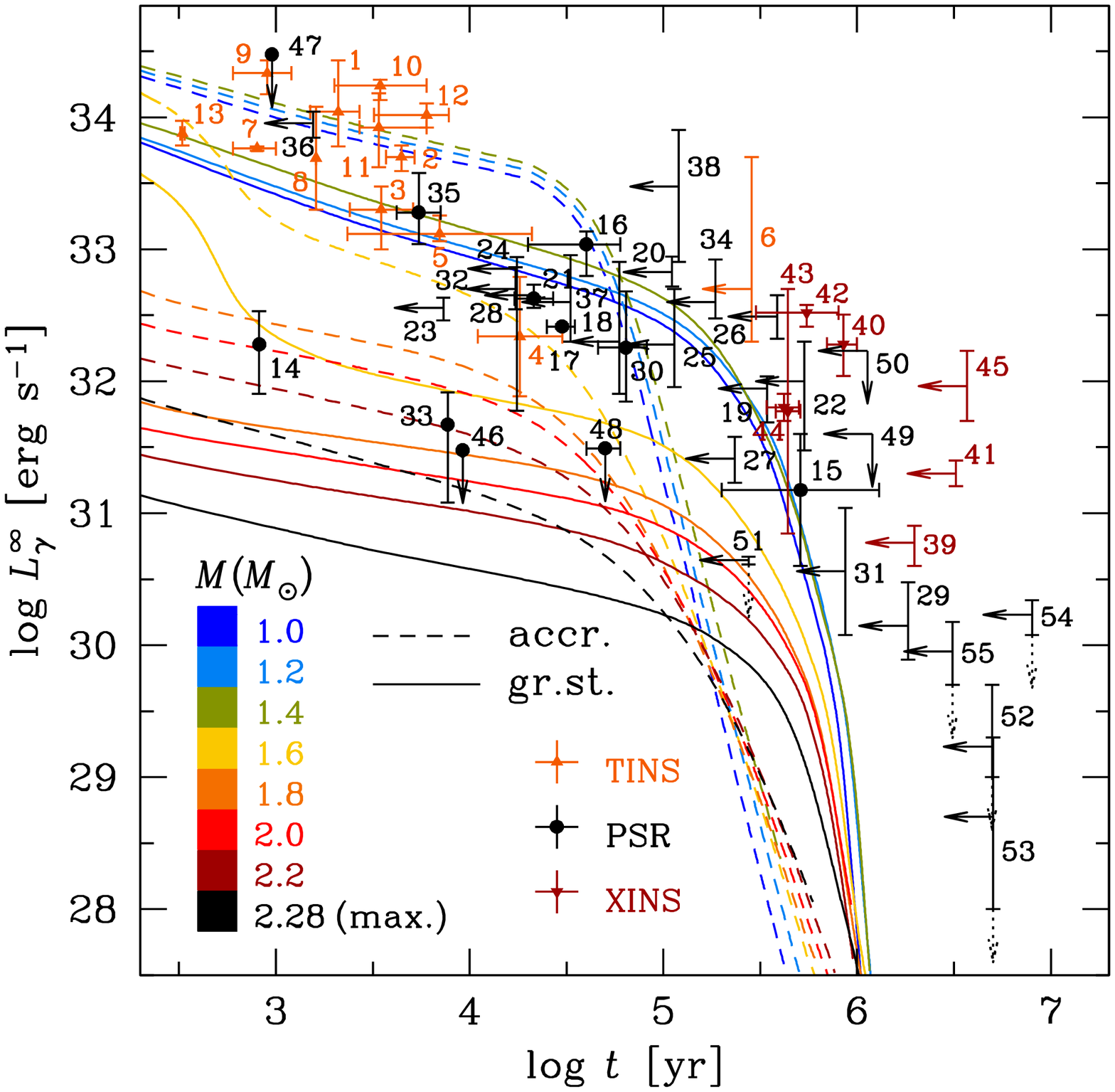}~
\includegraphics[width=.48\textwidth]{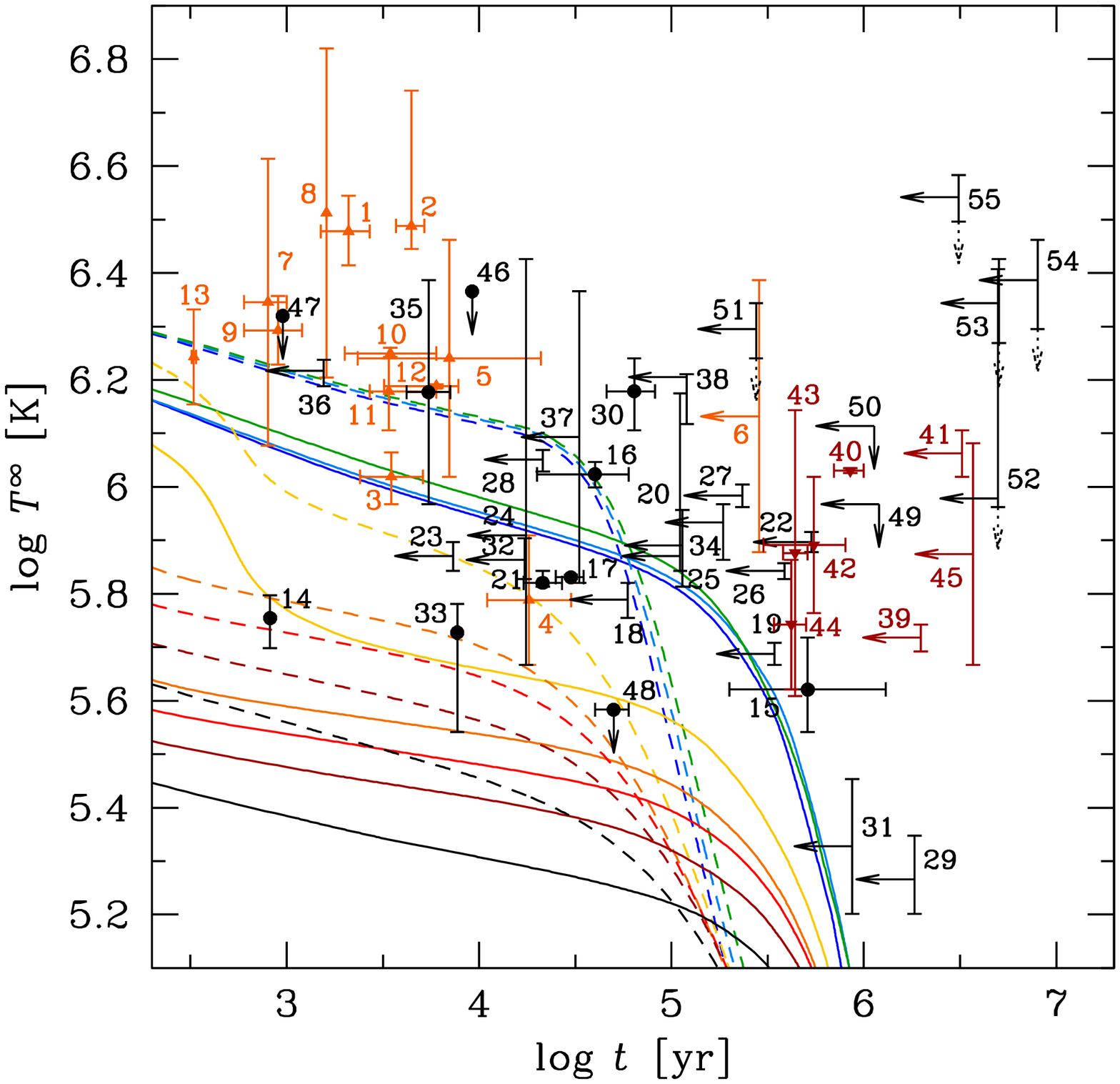}
\caption{\emph{Left panel}: Luminosities $\Ls$ plotted against ages $t$ from
Table~\ref{tab:AgeL} compared with theoretical cooling curves for
INSs with different masses (coded with color), non-accreted
(solid curves) and  accreted (dashed curves) blanketing envelopes, and with
the superfluidity model MSH+BS+TTav. The data for different
INS classes (TINSs, PSRs, XINSs) are plotted with the same
symbols and colors as in Fig.~\ref{fig:coolsfx}. Errorbars show
uncertainties and arrows show upper limits. If the detected thermal
radiation is thought to originate entirely from hot spots, vertical
errorbars are supplemented by broken downward arrows (meaning that the
non-detected thermal component can be fainter). Horizontal
errorbars show the estimated age intervals, whenever available;
otherwise
horizontal arrows mark less reliable characteristic ages. \emph{Right
panel}: the same for temperatures $\Teff^\infty$ instead of luminosities. See the text
for details.
\label{fig:coolmass}}
\end{figure*}

\section{Theory versus observations}
\label{sect:cooling}

Let us compare the cooling theory of INSs  with the observational data
described in Sect.~\ref{sec:cooldat}. We will demonstrate the effects of
heat blanketing envelopes, stellar mass and models of nucleon
superfluidity on the cooling curves. All these effects have been
described in the literature. Our aim is to attract attention to some
modern theoretical models and use the updated set of observational data.

To simulate neutron-star cooling, we use the numerical code
described in \citet{PC18}.
The physics input is mainly as
reviewed in \citet{PPP15}, supplemented by recent updates of the
superfluid pairing gaps \citep{Ding_16} and the modified Urca reaction
rates \citep{ShterninBH18}. The EoS and
the proton fraction in the core correspond to the BSk24 model \citep{Pearson_ea18}. 

\subsection{Effects of heat blanket for a star of fixed mass}
\label{sect:coolenv}

Let us first check the effects of heat blanketing envelopes
(Sect.~\ref{sect:blanket}) against a restricted data set of cooling INSs
with estimated ages. Fig.~\ref{fig:coolsfx} shows cooling curves
for a non-superfluid neutron star  of mass $M=1.4\,\msun$. The direct
Urca processes are forbidden in such a star (for the BSk24 model of
matter composition that we use), so that its neutrino cooling  is mostly
powered by the modified Urca processes. From a practical point of view,
it makes sense not to consider fast cooling without superfluidity,
because this would produce luminosities substantially below all
observations. The dashed curve in the figure shows the cooling of a
neutron star with ground-state crust, where the blanketing envelope 
consists of iron up to $\rho\approx8\times10^6$ \gcc{} and of nickel
isotopes at deeper layers. The solid curve shows the case where the
ground-state crust is replaced by carbon and then by oxygen up to the
densities of and temperatures of carbon and oxygen fusion, respectively
\citep{PC12}. The dot-dashed curve corresponds to the case where the
ground-state matter is replaced by helium at $\rho<10^9$ \gcc. The
shaded strips are formed by cooling curves calculated assuming different
possible amount of accreted material.

Points and errorbars in Fig.~\ref{fig:coolsfx} show redshifted
luminosities $L^\infty$ versus ages $t_*$. Errorbars
give uncertainties ($1\sigma$) of the measured values, and downward
arrows mark $3\sigma$ upper limits on $L^\infty$. 

According to  Fig.~\ref{fig:coolsfx}, variations of chemical composition in
the heat blanket of a 1.4\,\msun{} star allow one to explain much more objects, 
than in case of one star with fixed heat blanket, but not
all selected objects. The coldest stars at the neutrino cooling stage
have thermal
luminosities far below the theoretical curves (we will see
in Sect.~\ref{sect:coolmass} that this can be explained by
enhanced neutrino emission for INSs with large masses).

In contrast, 
the evolution of INSs at the photon cooling stage is not
regulated directly by their neutrino emission. However, INSs
observed at the photon cooling stage could not be 
very powerful neutrino emitters at the previous
stage. Otherwise, they would have lost too much heat 
and would now be too faint to be observed. As seen from
Fig.~\ref{fig:coolsfx},  these
non-superfluid INSs are reasonably compatible with the
standard neutrino cooling provided by the modified Urca processes. 

\subsection{Effects of fast cooling processes and superfluidity}
\label{sect:coolmass}

The proton fraction in the neutron star core  grows with $\rho$. In
central regions of the stars with  $M > M_\mathrm{DU}\sim1.6\,\msun$ (in
the BSk24 model), the proton fraction is above the threshold for opening
the powerful neutrino emission via direct Urca processes (see
Sect.~\ref{sect:regulators}). The higher $M$ above this threshold, the
larger the central part of the core where the direct Urca processes
operate. 

In Fig.~\ref{fig:coolmass} we compare theoretical luminosities and
surface temperatures of neutron stars of different masses with all data
in Table~\ref{tab:AgeL} (not only the objects with known $t_*$). When
available, we keep using $t_*$ for the age estimate. Otherwise we use
characteristic ages and treat them as upper limits (which are, however,
not strict, as discussed in Sect.~\ref{sect:descriptables}). If the
observed thermal luminosity is thought to be produced by hot spots, the
thermal flux from the interior must be smaller, and we supplement 
errorbars by dotted arrows directed downward. 

According to the theory, neutron star cooling can be greatly affected by
nucleon superfluidity (see Sect.~\ref{sect:regulators}). In
Fig.~\ref{fig:coolmass} we show cooling curves of superfluid neutron
stars of different masses. The stars are supposed to have either
non-accreted (ground state) heat blanketing envelopes or accreted
envelopes composed of helium and carbon. The critical temperatures for
singlet neutron, singlet proton and triplet neutron types of pairing as
functions of density are evaluated using the MSH, BS and TTav
parametrizations of \citet{Ho_ea15}. They are based on theoretical
models computed, respectively, by
\citet{MargueronSH08,BaldoSchulze07,TakatsukaTamagaki04}. For each
given type of the blanketing envelope, the cooling curves are close to
one another as long as $M<M_\mathrm{DU}$, but they become drastically
different at higher $M$.

Comparing the right  and left panels of Fig.~\ref{fig:coolmass}, we see
that if, instead of $L^\infty$, we employ observed surface temperatures
$T^\infty$ neglecting information on effective emitting sphere radius
$\Reff$, then  the agreement between the theory and the data becomes
generally worse and some upper limits become useless. This illustrates
the importance of luminosity estimates for the cooling theory (as was
previously stressed, e.g., by \citealt{Vigano_13}; also see the
discussion in Sect.~\ref{sect:gen}).

Comparing the data with the computed curves, we see that the models 
with iron blanketing envelopes are unable to explain the hottest of the
younger stars. Accreted envelopes improve the agreement, as we have already seen
in Sect.~\ref{sect:coolenv} for the particular case of $M=1.4\,\msun$ star.
The enhanced neutrino cooling through the direct Urca process for the
more massive stars allows us to explain low thermal
luminosities of some sources at the neutrino cooling stage, which
remained unexplained in Fig.~\ref{fig:coolsfx}.

The superfluid stars cool down faster at the photon cooling stage,
compared with their non-superfluid counterparts discussed in
Sect.~\ref{sect:coolenv}. The faster cooling is explained by the  heat
capacity $C(\Ti)$ reduced by superfluidity (see
Sect.~\ref{sect:regulators}).  As a consequence, according to 
Fig.~\ref{fig:coolmass}, some XINSs are significantly hotter than they
should have been at their estimated ages. One explanation may be that
some reheating is operating in these objects, but another possibility is
an overestimation of the critical temperature of neutrons in our cooling
models.

\begin{figure}
\includegraphics[width=.48\textwidth]{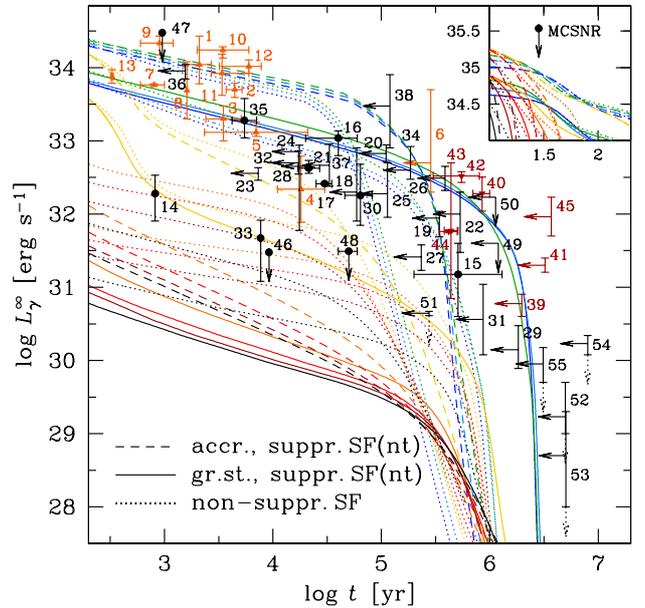}
\caption{The same as in Fig.~\ref{fig:coolmass},
but with suppressed triplet superfluid gap.
The cooling curves with non-suppressed gap
from Fig.~\ref{fig:coolmass}
are reproduced by dotted lines for comparison.
The inset shows the early cooling compared with the tentative upper
limit to the luminosity of a compact object in MCSNR J0535$-$6916
(the remnant of SN 1987A)
according to \citet{Cigan_99}.
\label{fig:cool36}}
\end{figure}

Recent studies have demonstrated that the effects of many-body
correlations on baryon superfluidity may strongly reduce the superfluid
gap for the triplet type of pairing (e.g., \citealp{Ding_16} and
references therein; see \citealt{SedrakianClark} for a discussion).  To
test an effect of such gap reduction on  neutron star cooling, we
multiply the TTav gap, used in Fig.~\ref{fig:coolmass}, by an
appropriate suppression factor. We evaluate this factor as the ratio of
the neutron triplet pairing gaps calculated with and without inclusion
of the many-body correlations \citep{Ding_16}. In this example we rely
on the results obtained with the Av18 effective potential that underlies
the TTav gap model. Whereas the TTav critical temperature 
$T_\mathrm{cn}$, as parametrized by \citet{Ho_ea15}, has a maximum of
$5.5\times10^8$~K at $\rho\approx4\times10^{14}$ \gcc, the reduced
$T_\mathrm{cn}$ barely reaches $10^8$~K at
$\rho\approx3\times10^{14}$ \gcc{} and falls below $10^7$~K at
$\rho\gtrsim6\times10^{14}$ \gcc. The resulting cooling curves are shown
in Fig.~\ref{fig:cool36}. The smaller pairing gap cannot strongly reduce
the heat capacity of neutrons in  the star cores. Moreover, the neutrons
remain normal (non-superfluid) in a substantial  range of densities in
the core. The heat capacity remains relatively large, and the
temperature decreases slower. Thus our calculations demonstrate that
thermal luminosities of rather warm and old cooling INSs strongly depend
on the neutron superfluidity in their cores. The agreement with
observations of relatively old cooling neutron stars improves
substantially by reduction of the neutron triplet pairing gap. This
confirms the analogous conclusions, recently obtained by other authors
based on a smaller sample of cooling neutron stars
\citep{Taranto_16,Fortin_18,Beznogov_18,Wei_19,Wei_20}, and also by
analysis of thermal luminosities of soft X-ray
transients in quiescence \citep{Fortin_18,PCC19}.

The inset of Fig.~\ref{fig:cool36} shows early cooling
($10\mbox{~yr}<t<200$~yr and $L^\infty>10^{34}$ erg s$^{-1}$) compared
with a tentative upper limit to the luminosity of a possible compact
object in the supernova remnant MCSNR J0535$-$6916, left after the SN
1987A explosion. This limit, $L^\infty<90\,L_\odot$ at $t=28.5$~yr, 
was obtained by
\citet{Cigan_99} by an analysis of high resolution ALMA images of dust
and molecules in the SN 1987A ejecta; implications of this possible
detection for the neutron star cooling theory were discussed by
\citet{Page_20}.

Our analysis is illustrative and naturally incomplete. For instance, we
have used  one model EoS. Other EoSs give different cooling curves; in
particular, different models predict different masses $M_{\rm DU}$ for
opening fast neutrino cooling. We have also used a limited set of proton
and neutron critical temperature profiles $T_{\rm cn}(\rho)$ and $T_{\rm
cp}(\rho)$ which are extremely model dependent and which can strongly
affect the cooling curves.

\section{Conclusions}
\label{sect:concl}

We have revised available observational estimates of ages and thermal
luminosities of middle-aged cooling isolated neutron stars and composed
a catalogue of their key observational properties for testing
theoretical models. A comparison with simulations based on several
cooling scenarios shows that the theory can be in good agreement with
the data.  The estimates of thermal luminosities are usually in better
agreement with the cooling theory than temperature estimates for some
neutron stars (which may be related to non-uniform temperature
distributions over the surfaces of these stars, as  discussed in 
Sect.~\ref{sect:gen}). 

The data suggest that the enhanced neutrino emission due to direct Urca
reactions operate in some (but not in all) neutron stars, which prompts
that these neutron stars are more massive than other, so that the proton
fraction in their interiors reaches the values,  sufficient to enable
the direct Urca processes.  The data also favour suppression of the
neutron triplet superfluidity, in agreement with recent
analyses of smaller observational data
sets.

\section*{acknowledgements}

We thank the referee, Dr.\ Andreas Reisenegger, for many useful remarks
and comments, which helped us to improve the paper. The work of A.P.,
D.Z.{} and Yu.S.{} was partially funded by RFBR and DFG according to the
research project 19-52-12013. D.Y.{} was partially supported by the
grant 14.W03.31.0021 of the Ministry of Science and Higher Education of
the Russian Federation. M.B.{} is partially supported by the Consejo
Nacional de Ciencia y Technolog\'{\i}a with a CB 2014-1 grant \#240512.
M.B.{} also acknowledges support from a postdoctoral fellowship from
UNAM DGPAPA.

\section*{Data availability}

We present the collected data on a dedicated Web
page \url{http://www.ioffe.ru/astro/NSG/thermal/cooldat.html};
we are planning to update this page on regular basis as new
observational data on cooling neutron stars appear.

\bibliographystyle{mnras}

%


\bibliography{cooldat}

\onecolumn

\begin{center}
\begin{longtable}{r l c  c  c  c @{} c}
  \caption{Isolated middle-aged cooling neutron stars.} 
  \label{tab:CoolList} \\
\hline
\hline
 no.  & Identifier           &  Association     &  $\mathcal{P}$   & $B_\mathrm{dip}$ & Distance & Refs.  \\
      &                      &  or nickname     &  (s)   &       ($10^{12}$\,G)       &   (kpc)  &        \\
\hline
  \endfirsthead
  \multicolumn{7}{c}
  {{\tablename\ \thetable{} -- \emph{continued}}} \\
\hline
\hline
 no.  & Identifier           &  Association     &  $\mathcal{P}$   & $B_\mathrm{dip}$ & Distance & Refs.  \\
      &                      &  or nickname     &  (s)   &       ($10^{12}$\,G)       &   (kpc)  &        \\
\hline
  \endhead
\hline
 \multicolumn{7}{r}{\emph{continued on next page}}
  \endfoot
  \hline
\hline
  \endlastfoot
\multicolumn{7}{c}{I. Weakly magnetized thermal emitters}\\
 1 & 1E 0102.2$-$7219        & B0102$-$72.3 in SMC & --- &    ---  & $62\pm2$                  & 1(d)         \\[.2ex]
 2 & RX J0822.0$-$4300       & Puppis A         & 0.113  &   0.029 & $2.2\pm0.3$               & 2(d)         \\[.2ex]
 3 & CXOU J085201.4$-$461753 & Vela Jr.         &  ---   &    ---  & $0.77\pm0.2$              & 3(d)         \\[.2ex]
 4 & 2XMM J104608.7$-$594306 & Homunculus       &  ---   &    ---  & $2.35\pm1 $               & 4(d)         \\[.2ex]
 5 & 1E 1207.4$-$5209        & G296.5+10.0      & 0.424  &   0.098 & $2.1^{+1.8}_{-0.8}$       & 5(t), 6(d)   \\[.2ex]
 6 & 1RXS J141256.0+792204   &  `Calvera'       & 0.0592 &   0.45  & $\sim1$\,--\,4            & 7--9(t,d)    \\[.2ex]
 7 & CXOU J160103.1$-$513353 & G330.2+1.0       &  ---   &    ---  & $4.9^{+5}_{-0.3}$         & 10(d)        \\[.2ex]
 8 & 1WGA J1713.4$-$3949     & G347.3$-$0.5     &  ---   &    ---  & $1.3\pm0.4$               & 11(d)        \\[.2ex]
 9 & XMMU J172054.5-372652   & G350.1$-$0.3     &  ---   &    ---  & $6.1^{+1.3}_{-1.0}$       & 12(d)        \\[.2ex]
10 & XMMU J173203.3$-$344518 & HESS J1731$-$347 &  ---   &    ---  & $3.2\pm0.8$               & 13,14(d)     \\[.2ex]
11 & CXOU J181852.0$-$150213 & G015.9+00.2      &  ---   &    ---  & $10.0^{+6.7}_{-1.5}$      & 15(d)        \\[.2ex]
12 & CXOU J185238.6+004020   & Kes 79           & 0.105  &   0.031 & $7.1^{+0.4}_{-0.6}$       & 16(d)        \\[.2ex]
13 & CXOU J232327.8+584842   & Cas A            &  ---   &    ---  & $3.4^{+0.3}_{-0.1}$       & 17(d)        \\[.2ex]
 \multicolumn{7}{c}{II. Ordinary pulsars}\\[.2ex]
14 & PSR J0205+6449          &   3C 58          & 0.0657 &  3.6    & $2.0\pm0.3$               & 18(d)        \\[.2ex]
15 & PSR J0357+3205          &  `Morla'         & 0.444  &  2.4    & $0.45\pm0.05$             & 19(d)        \\[.2ex]
16 & PSR J0538+2817          & Sim 147          & 0.143  &  0.73   & $1.47^{+0.42}_{-0.27}$    & 20(d)        \\[.2ex]
17 & CXOU J061705.3+222127   & IC 443           & $0.25^{+0.35}_{-0.15}$  & 4\,--\,40  & $1.7\pm0.3$ & 21(t), 22\,--\,24(d) \\[.2ex]
18 & PSR J0633+0632          & Collinder 106 (?)& 0.297  &  4.9    & $0.9^{+1.1}_{-0.1}$       & 25(t,d)      \\[.2ex]
19 & PSR J0633+1746          & `Geminga'        & 0.237  &  1.6    & $0.25^{+0.23}_{-0.08}$    & 26(d)        \\[.2ex]
20 & PSR B0656+14            & Monogem Ring     & 0.385  &  4.7    & $0.28\pm0.03$             & 26(d)        \\[.2ex]
21 & PSR B0833$-$45          & Vela             & 0.0893 &  3.4    & $0.28\pm0.02$             & 26(d)        \\[.2ex]
22 & PSR B1055$-$52          &    ---           & 0.197  &  1.1    & $0.35\pm0.15$             & 27(d)        \\[.2ex]
23 & PSR J1357$-$6429        & HESS J1356$-$645 & 0.166  &  7.8    & $\sim2.5$                 & 28(d)        \\[.2ex]
24 & PSR B1706$-$44          & G343.1$-$02.3    & 0.102  &  3.1    & $2.6^{+0.5}_{-0.6}$       & 26(d)        \\[.2ex]
25 & PSR J1740+1000          &    ---           & 0.154  &  1.8    & $\sim1.2$\,--\,1.4        & 29(d)        \\[.2ex]
26 & PSR J1741$-$2054        &    ---           & 0.414  &  2.7    & $0.8\pm0.3$               & 30(d)        \\[.2ex]
27 & PSR B1822$-$09          &    ---           & 0.769  &  6.4    & 0.9\,--\,1.9              & 31(d)        \\[.2ex]
28 & PSR B1823$-$13          &    ---           & 0.101  &  2.8    & $4\pm1$                   & 32(d)        \\[.2ex]
29 & PSR J1836+5925          & `Next Geminga'   & 0.173  &  0.52   & $\sim0.2-0.7$             & 33,34(d)     \\[.2ex]
30 & PSR B1951+32            &   CTB 80         & 0.0395 &  0.49   & $2.4\pm0.2$               & 35(d)        \\[.2ex]
31 & PSR J1957+5033          &    ---           & 0.375  &  1.6    &  $\sim0.1-0.8$            &36(t,d), 37(d)\\[.2ex]
32 & PSR J2021+3651          &    ---           & 0.104  &  3.2    & $1.8^{+1.7}_{-1.4}$       & 38(d)        \\[.2ex]
33 & PSR B2334+61            & G114.3+00.3      & 0.495  &  9.9    &  $\sim0.7$                & 39(d)        \\[.2ex]
 \multicolumn{7}{c}{III. High-B pulsars}\\[.2ex]
34 & PSR J0726$-$2612        &    ---           & 3.442  &  32     & $1.0^{+1.4}_{-0.7}$       & 40,41(d)     \\[.2ex]
35 & PSR J1119$-$6127        & G292.2$-$00.5    & 0.408  &  41     & $8.4\pm0.4$               & 42(d)        \\[.2ex]
36 & PSR B1509$-$58          &    ---           & 0.151  &  15     & $5.2\pm1.4$               & 43(d)        \\[.2ex]
37 & PSR J1718$-$3718        &    ---           & 3.379  &  75     & 4.5\,--\,10               & 44(d)        \\[.2ex]
38 & PSR J1819$-$1458        &    ---           & 4.263  &  50     & $3.6\pm0.9$               & 45(d)        \\[.2ex]
 \multicolumn{7}{c}{IV. The Magnificent Seven}\\[.2ex]
39 & RX J0420.0$-$5022       &    ---           & 3.453  &  9.9    &  0.325\,--\,0.345         & 46(t), 47(d) \\[.2ex]
40 & RX J0720.4$-$3125       &    ---           & 16.782 &  24     & $0.286^{+27}_{-23}      $ &48(t), 47,49(d)\\[.2ex]
41 & RX J0806.4-4123         &    ---           & 11.370 &  26     &  0.235\,--\,0.250         & 47(d)        \\[.2ex]
42 & RX J1308.6+2127         &    ---           & 10.312 &  34     & $0.38^{+0.02}_{-0.03}   $ & 50,51(d)     \\[.2ex]
43 & RX J1605.3+3249         &    ---           &  ---   &   ---   &  0.09\,--\,0.4            & 52,53(t), 47,52--55(d) \\[.2ex]
44 & RX J1856.5$-$3754       &  Upper Scorpius  & 7.055  &  15     & $0.123^{+0.011}_{-0.015}$ & 56(t), 57(d) \\[.2ex]
45 & RX J2143.0+0654         &    ---           & 9.428  &  200    &  0.390\,--\,0.430         & 58(t), 47(d) \\[.2ex]
 \multicolumn{7}{c}{V. Neutron stars with upper limits on thermal emission}\\[.2ex]
46 & PSR J0007+7303          &    CTA~1         & 0.316  &  1.1    &    $1.4\pm0.3$            & 59(d)        \\[.2ex]
47 & PSR B0531+21            &   Crab           & 0.0334 &  3.8    & $2.0^{+0.7}_{-0.6}$       & 60(d)        \\[.2ex]
48 & PSR B1727$-$47          &   RCW 114        & 0.830  &  12     &  0.5\,--\,0.8             & 61,62(d)     \\[.2ex]
49 & PSR J2043+2740          &  Cygnus Loop (?) &0.0961  &  0.35   &   1.5--1.8                & 63,64(d)     \\[.2ex]
50 & PSR B2224+65            &  Guitar          & 0.683  &  2.6    & $0.83^{+0.17}_{-0.10}$    & 65(d)        \\[.2ex]
 \multicolumn{7}{c}{VI. Middle-aged pulsars with measured thermal emission of hot spots}\\[.2ex]
51 & PSR B0114+58            &    ---           & 0.101  &  0.78   &  $1.8\pm0.6$              & 66(d)        \\[.2ex]
52 & PSR B0943+10            &    ---           & 1.098  &  2.0    &  0.63\,--\,0.89           & 67(d)        \\[.2ex]
53 & PSR B1133+16            &    ---           & 1.188  &  2.1    & $0.35\pm0.02         $    & 26(d)        \\[.2ex]
54 & PSR J1154-6250          &    ---           & 0.282  &  0.40   &  1.36\,--\,1.77           & 68(d)        \\[.2ex]
55 & PSR B1929+10            &    ---           & 0.227  &  0.52   & $0.31^{+0.09}_{-0.05}$    & 26(d)        \\[.2ex]
\hline
\multicolumn{7}{@{}p{\linewidth}}{\rule{0pt}{3ex}
\textbf{References:} 
 1.~\citet{SMCdistance};
 2.~\citet{Reynoso_ea03};
 3.~\citet{Allen2015-velajr};
 4.~\citet{Smith06};
 5.~\citet{HalpernGotthelf15};
 6.~\citet{Giacani_ea00};
 7.~\citet{HalpernBG13};
 8.~\citet{Zane_ea11};
 9.~\citet{Shibanov_16};
10.~\citet{McClureGriffiths_01};
11.~\citet{CassamChenai_04};
12.~{This work (spectral fit)};
13.~\citet{Klochkov_etal15};
14.~\citet{Maxted_ea18};
15.~\citet{Sasaki_ea18};
16.~\citet{Giacani_ea09};
17.~\citet{Reed_ea95};
18.~\citet{Kothes13};
19.~\citet{Kirichenko_etal14};
20.~\citet{Ng_ea07};
21.~\citet{Swartz2015-ic443};
22.~\citet{Fesen84};
23.~\citet{WelshSallmen03};
24.~\citet{KochanekAB19};
25.~\citet{Danilenko_J0633};
26.~\citet{Verbiest_ea12};
27.~\citet{MignaniPK10};
28.~\citet{Zavlin07};
29.~\citet{Kargaltsev_ea12};
30.~\citet{Karpova_etal14};
31.~\citet{Hermsen2017PSRB1822-09};
32.~\citet{PavlovKB08};
33.~\citet{Abdo_10_psrJ1836};
34.~\citet{Arumugasamy2015PhD};
35.~\citet{StromStappers00};
36.~\citet{Marelli_15};
37.~\citet{Zyuzin_J1957};
38.~\citet{Kirichenko_15};
39.~\citet{Yar-Uyaniker2004-psrb2334};
40.~\citet{SpeagleKVK11};
41.~\citet{Rigoselli_J0726};
42.~\citet{Caswell_ea04};
43.~\citet{Gaensler_ea99};
44.~\citet{Zhu_ea11};
45.~\citet{Miller_ea13};
46.~\citet{KvK11};
47.~\citet{Posselt_ea07};
48.~\citet{Hambaryan_17};
49.~\citet{Tetzlaff_ea11};
50.~\citet{Motch_ea09};
51.~\citet{Hambaryan_ea11};
52.~\citet{Pires_ea19};
53.~\citet{Malacaria_19};
54.\citet{Motch_05};
55.~\citet{Tetzlaff_ea12};
56.~\citet{vKK08};
57.~\citet{Walter_ea10};
58.~\citet{KvK09};
59.~\citet{Pineault_93};
60.~\citet{Trimble73};
61.~\citet{Shternin_ea19};
62.~\citet{Zyuzin_B1727};
63.~\citet{Becker_04};
64.~\citet{Testa_18};
65.~\citet{Deller_19};
66.~\citet{RigoselliMereghetti18};
67.~\citet{Rigoselli_B0943};
68.~\citet{Igoshev_ea18}.
}
\end{longtable}

\end{center}


\begin{longtable}{r l c c c c @{\!} c}
 \caption{Ages and thermal radiation of cooling neutron stars.
 \label{tab:AgeL}} \\
 \hline
\hline
 no.  & Short   &  $t_\mathrm{c}$     &$t_*$  & $L^\infty$           & ${\kB}T^\infty$        & Refs. \\
      &  name   &      (kyr)          &  (kyr)         & ($10^{32}$ erg s$^{-1}$)    &    (eV)                &       \\
\hline
  \endfirsthead
  \multicolumn{7}{c}
  {{\tablename\ \thetable{} -- \emph{continued}}} \\
  \hline
\hline
 no.  & Short   &  $t_\mathrm{c}$     &$t_*$  & $L^\infty$           & ${\kB}T^\infty$        & Refs. \\
      &  name   &      (kyr)          &  (kyr)         & ($10^{32}$ erg s$^{-1}$)    &    (eV)                &       \\
\hline
   \endhead
  \hline
 \multicolumn{7}{r}{\emph{continued on next page}}
  \endfoot
  \hline
\hline
  \endlastfoot
 \multicolumn{7}{c}{I. Weakly magnetized thermal emitters}\\
 1 & 1E 0102 &   ---            & $2.1\pm0.6$          & $110^{+160}_{-50}$          & $210^{+40}_{-30}$      & 1(a), 2(s)       \\[.2ex]
 2 & Puppis A NS & $2.54\times10^5$ & $4.45\pm0.75$    & $50\pm11$                   & $276\pm15/455\pm20$    & 3(a), 4(s)       \\[.2ex]
 3 & Vela Jr.\ NS &   ---       & 2.1\,--\,5.4         & $20\pm10$                   & $90\pm10$              & 5(a), 6(s)       \\[.2ex]
 4 & J1046   &   ---            & 11\,--\,30           & $0.8-6$                     & 40\,--\,70             & 7(a,s)           \\[.2ex]
 5 & 1E 1207 & $3.01\times10^5$ & $7^{+14}_{-5}$       & $13.1^{+4.9}_{-1.6}$        & 90\,--\,250            & 8(a), 9(s)       \\[.2ex]
 6 & Calvera & $285$            &   ---                & 2\,--\,50                   & 65\,--\,210            & 10,11(s)         \\[.2ex]
 7 & J1601   &   ---            & $0.8\pm0.2$          & $58\pm2$                    & $118\pm1$              & 12(a), 13(s)     \\[.2ex]
 8 & J1713   &   ---            &  1.608               & $\sim20-120$                & $138\pm1$              & 14(a), 15(s)     \\[.2ex]
 9 & J1720   &   ---            &  0.6\,--\,1.2        & 150\,--\,270                & $161\pm9$              & 16(a), 15(s)     \\[.2ex]
10 & J1732   &   ---            &  2\,--\,6            & $174^{+19}_{-39}$           & $153^{+4}_{-2}$        & 17(a), 18(s)     \\[.2ex]
11 & J1818   &   ---            & $3.4^{+2.6}_{-0.7}$  & $84^{+68}_{-42}$            & $130\pm20$             & 19(a), 20(s)     \\[.2ex]
12 &Kes 79 NS& $1.92\times10^5$ & $6.0^{+1.8}_{-2.8}$  & $104^{+24}_{-20}$           & $133\pm1$              & 21(a), 22(s)     \\[.2ex]
13 & Cas A NS&   ---            & 0.320\,--\,0.338     &  61\,--\,94                 & 123\,--\,185           &23(a), 24\,--\,26(s)\\[.2ex]
 \multicolumn{7}{c}{II. Ordinary pulsars}\\[.2ex]
14 & J0205   & $5.37$           &    0.819             & $1.9^{+1.5}_{-1.1}$         &   $49^{+5}_{-6}$       & 27,28(a), 15(s)  \\[.2ex]
15 & Morla   & $541$            & 200\,--\,1300        & $0.15^{+0.25}_{-0.11}$      & $36^{+9}_{-6}$         & 29(a,s)          \\[.2ex]
16 & J0538   & $620$            & $40\pm20$            & $10.9^{+2.7}_{-4.6}$        & $91\pm5$               & 30(a), 31(a,s)   \\[.2ex]
17 & J0617   & 10\,--\,100      &  $\sim30$            & $2.6\pm0.1$                 & $58.4^{+0.6}_{-0.4}$   & 32(a), 33(s)     \\[.2ex]
18 & J0633   & $59.2$           &   ---                & $1.5^{+2.5}_{-0.9}$         & $53\pm4$               & 34(s)            \\[.2ex]
19 & Geminga & $342$            &   ---                &$0.88^{+0.21}_{-0.39}$       & $42\pm2$               & 35,36(s)         \\[.2ex]
20 & B0656   & $111$            &   ---                & $6.7^{+2.1}_{-1.5}$         & $64\pm4/123^{+6}_{-5}$ & 37(s)            \\[.2ex]
21 & Vela pulsar & $11.3$       & 17\,--\,23           & $4.24\pm0.12$               & $57^{+3}_{-1}$         & 38(a), 39(s)     \\[.2ex]
22 & B1055   & $535$            &   ---                & $1.0^{+1.0}_{-0.7}$         & $68\pm3$               & 40(s)            \\[.2ex]
23 & J1357   & $7.31$           &   ---                & $3.6\pm0.7$                 & $64\pm4$               & 41,42(s)         \\[.2ex]
24 & B1706   & $17.5$           &   ---                & $7.1^{+1.6}_{-6.5}$         & $71^{+140}_{-30}$      & 43(s)            \\[.2ex]
25 & J1740   & $114$            &   ---                & $1.9^{+3.1}_{-1.0}$         & $67\pm11$              & 44(s)            \\[.2ex]
26 & J1741   & $386$            &   ---                & $3.1^{+1.4}_{-1.0}$         & $60\pm2$               & 45(s)            \\[.2ex]
27 & B1822   & $233$            &   ---                & $0.26^{+0.12}_{-0.09}$      & $83\pm4$               & 46(s)            \\[.2ex]
28 & B1823   & $21.4$           &   ---                & $4.5\pm0.9$                 & $97^{+4}_{-5}$         & 47(s)            \\[.2ex]
29 & Next Geminga & $1.83\times10^3$ &   ---           & $0.014^{+0.016}_{-0.006}$   & $15.9^{+3.3}_{-2.2}$   & 48(s)            \\[.2ex]
30 & B1951   & $107$            &   $64\pm18$          & $1.8^{+3.0}_{-1.1}$         & $130\pm20$             & 49(a), 50(s)     \\[.2ex]
31 & J1957   & $870$            &     ---              & 0.012\,--\,0.11             & $\sim13$\,--\,25       & 51(s)            \\[.2ex]
32 & J2021   & $17.2$           &   ---                &       $5^{+3}_{-2}$         & $63^{+6}_{-5}$         & 52(s)            \\[.2ex]
33 & B2334   & $40.6$           &   $\sim7.7$          & $0.47\pm0.35$               & $38^{+6}_{-9}$         & 53(a), 54(s)     \\[.2ex]
 \multicolumn{7}{c}{III. High-B pulsars}\\[.2ex]
34 & J0726   & $186$            &   ---                & $4.0^{+4.4}_{-1.0}$         & $74^{+6}_{-11}$        & 55(s)            \\[.2ex]
35 & J1119   & $1.61$           & 4.2\,--\,7.1         & $19^{+19}_{-8}$             & $\sim80$\,--\,210      & 56(a), 57(s)     \\[.2ex]
36 & B1509   & $1.56$           &   ---                & $90\pm20$                   & $142^{+7}_{-9}$        & 58(s)            \\[.2ex]
37 & J1718   & $33.2$           &   ---                & $4^{+5}_{-2}$               & 57\,--\,200            & 59(s)            \\[.2ex]
38 & J1819   & $120$            &   ---                & $30^{+50}_{-22}$            & $138^{+3}_{-25}$       & 60(s)            \\[.2ex]
 \multicolumn{7}{c}{IV. The Magnificent Seven}\\[.2ex]
39 & J0420   & $1.98\times10^3$ &   ---                & $0.06\pm0.02$               & $45.0\pm2.6$           & 61,62(s)         \\[.2ex]
40 & J0720   & $1.90\times10^3$ & $850\pm150$          & $1.9^{+1.3}_{-0.8}$         & 90\,--\,100            & 63(a), 64,65(s)  \\[.2ex]
41 & J0806   & $3.24\times10^3$ &   ---                & 0.16\,--\,0.25              & $\sim90-110$           & 61,62(s)         \\[.2ex]
42 & J1308   & $1.46\times10^3$ & $550\pm250$          & $3.3^{+0.5}_{-0.7}$         & $\sim50$\,--\,90       & 66(a), 67(a,s)   \\[.2ex]
43 & J1605   &    ---           & $440^{+70}_{-60}$    & 0.07\,--\,5                 & 35\,--\,120            & 68(a), 69,70(s)  \\[.2ex]
44 & J1856   & $3.76\times10^3$ & $420\pm80$           & 0.5\,--\,0.8                & 36\,--\,63             &71(a), 72\,--74(s)\\[.2ex]
45 & J2143   & $3.7\times10^3$  &   ---                & 0.5\,--\,1.7                & 40/100                 & 75(s)            \\[.2ex]
 \multicolumn{7}{c}{V. Upper limits}\\[.2ex]
46 & J0007   & $13.9$           &    $\approx9.2$      & $<0.3$                      &   $<200$               & 76(a), 77(s)     \\[.2ex]
47 & Crab pulsar& $1.26$        &    0.954             & $<300$                      &   $<180$               & 78(a), 79(s)     \\[.2ex]
48 & B1727   & $80.5$           &  $50\pm10$           & $<0.35$                     &   $<33$                & 80(a), 81(s)     \\[.2ex]
49 & J2043   & $1.20\times10^3$ &     ---              & $<0.4$                      &   $<80$                & 82\,--\,84(s)    \\[.2ex]
50 & Guitar pulsar & $1.13\times10^3$ &     ---              & $<1.7$                      &   $<110$               & 85(s)            \\[.2ex]
 \multicolumn{7}{c}{VI. Hot spots}\\[.2ex]
51 & B0114   & $275$            &   ---                & $0.044\pm0.003$             & $170\pm20$             & 86(s)            \\[.2ex]
52 & B0943   & $4.98\times10^3$ &   ---                & 0.001\,--\,0.005            &$82^{+3}_{-9}-(\sim220)$& 87(s)            \\[.2ex]
53 & B1133   & $5.04\times10^3$ &   ---                & $0.0003^{+0.0017}_{-0.0002}$& $190^{+40}_{-30}$      & 88(s)            \\[.2ex]
54 & J1154   & $7.99\times10^3$ &   ---                & $0.017\pm0.05$              & $210\pm40$             & 89(s)            \\[.2ex]
55 & B1929   & $3.11\times10^3$ &   ---                & $0.0084^{+0.0034}_{-0.0022}$& $300^{+20}_{-30}$      & 90(s)            \\[.2ex]
\hline
\multicolumn{7}{@{}p{\linewidth}}{\rule{0pt}{3ex}
\textbf{References:} 
 1.~\citet{Xi_19};
 2.~\citet{HebbarHH20};
 3.~\citet{Becker_ea12};
 4.~\citet{DeLuca_ea12};
 5.~\citet{Allen2015-velajr};
 6.~\citet{Danilenko_15};
 7.~\citet{Pires_ea15};
 8.~\citet{Roger_ea88};
 9.~\citet{Mereghetti_ea02};
10.~\citet{Shibanov_16};
11.~\citet{Bogdanov_ea19};
12.~\citet{Borkowski2018-age-cco-snrg330};
13.~\citet{Doroshenko2018-cco-snrg330};
14.~\citet{CassamChenai_04};
15.~{This work};
16.~\citet{Lovchinsky_11};
17.~\citet{CuiPS16};
18.~\citet{Klochkov_etal15};
19.~\citet{Sasaki_ea18};
20.~\citet{Klochkov_ea16};
21.~\citet{Sun_ea04};
22.~\citet{Bogdanov14};
23.~Flamsteed (1680), \citet{Ashworth80};
24.~\citet{HH10};
25.~\citet{PosseltPavlov18};
26.~\citet{Wijngaarden_19};
27.~\citet{Stephenson71};
28.~\citet{Kothes13};
29.~\citet{Kirichenko_etal14};
30.~\citet{Kramer_03};
31.~\citet{Ng_ea07};
32.~\citet{Chevalier1999};
33.~\citet{Swartz2015-ic443};
34.~\citet{Danilenko_J0633};
35.~\citet{DeLuca_ea05};
36.~\citet{Mori_ea14};
37.~\citet{Arumugasamy_ea18};
38.~\citet{Aschenbach02};
39.~\citet{2018VELA};
40.~\citet{MignaniPK10};
41.~\citet{Zavlin07};
42.~\citet{Chang_ea12};
43.~\citet{McGowan_etal04};
44.~\citet{Kargaltsev_ea12};
45.~\citet{Karpova_etal14};
46.~\citet{Hermsen2017PSRB1822-09};
47.~\citet{PavlovKB08};
48.~\citet{Arumugasamy2015PhD};
49.~\citet{Migliazzo_02};
50.~\citet{Li2005-psrb1951};
51.~\citet{Zyuzin_J1957};
52.~\citet{Kirichenko_15};
53.~\citet{Yar-Uyaniker2004-psrb2334};
54.~\citet{McGowan_ea06};
55.~\citet{Rigoselli_J0726};
56.~\citet{KumarSHG12};
57.~\citet{Ng_ea12};
58.~\citet{Hu_ea17};
59.~\citet{Zhu_ea11};
60.~\citet{Miller_ea13};
61.~\citet{Haberl_ea04};
62.~\citet{KvK09};
63.~\citet{Tetzlaff_ea11};
64.~\citet{Hohle_ea12};
65.~\citet{Hambaryan_17};
66.~\citet{Motch_ea09};
67.~\citet{Hambaryan_ea11};
68.~\citet{Tetzlaff_ea12};
69.~\citet{Pires_ea19};
70.~\citet{Malacaria_19};
71.~\citet{Mignani_ea13};
72.~\citet{Ho_etal07};
73.~\citet{Sartore_ea12};
74.~\citet{Yoneyama_ea17};
75.~\citet{Schwope_ea09};
76.~\citet{MartinTP16};
77.~\citet{Caraveo2010-psrj0007};
78.~\citet{StephensonGreen03};
79.~\citet{Weisskopf_ea11};
80.~\citet{Shternin_ea19};
81.~\citet{Zyuzin_B1727};
82.~\citet{Becker_04};
83.~\citet{ZavlinPavlov04};
84.~\citet{Zavlin09};
85.~\citet{HuiBecker07};
86.~\citet{RigoselliMereghetti18};
87.~\citet{Rigoselli_B0943};
88.~\citet{Szary_ea17};
89.~\citet{Igoshev_ea18};
90.~\citet{MisanovicPG08}.
}
\end{longtable}

\label{lastpage}
\end{document}